\documentclass[preprint,12pt]{elsarticle}

\usepackage{color}
\usepackage{soul}
\usepackage{amssymb}
\usepackage{lipsum}
\usepackage{hyperref}
\usepackage{multirow}
\usepackage[table]{xcolor}

\usepackage{tcolorbox}
\usepackage{booktabs}
\usepackage{amsmath}
\usepackage{graphicx} 
\usepackage{subcaption} 

\journal{}

\begin{document}

\begin{frontmatter}



\title{Evaluating Quantum Kernel Methods for Track-Based Classification in High-Energy Physics}

\corref{*}
\author[a]{Emmanuel Billias 
\corref{corresponding-author}
}

\author[a,b]{Nikos Chrisochoides
}

\address[a]{CRTC, Department of Computer Science, Old Dominion University, Norfolk, VA, USA}
\address[b]{Department of Physics, Old Dominion University, Norfolk, VA, USA}

\cortext[corresponding-author]{Corresponding author. \textit{E-mail address:} ebill007@odu.edu}



\begin{abstract}
We present a systematic design for large-scale quantum kernel classification, demonstrated through a quantum support vector classifier (QSVC) for particle–track classification using centroid-based CLAS12 drift-chamber features. Each event is encoded into a six-qubit state via a fully entangled \texttt{ZZFeatureMap}, whose fidelities define a quantum kernel within a standard SVM framework. By decoupling state preparation from kernel construction and distributing evaluation across a multi-node MPI-based HPC allocation, the approach scales to $1.0\times10^5$ training and $4.0\times10^5$ test events with an exactly constructed kernel matrix, to our knowledge more than an order of magnitude larger than prior high-energy-physics quantum-kernel studies.

Benchmarked against linear, polynomial, RBF, and sigmoid SVM kernels and extremely randomized trees (ERT), the ideal QSVC achieves the highest recall (99.99\%) among all models. Under a calibrated hardware noise model (FakeMumbaiV2, 500 training / 2,000 test events), AUC falls from 0.9985 to 0.9671 and peak significance improvement falls from 17.5 to $\sim$3.5, yet recall remains at 99.51\% — indicating this signal-retention advantage is attenuated but not eliminated by circuit-level decoherence.

Geometric analysis of the quantum embedding shows near-orthogonal inter-class states with coherent intra-class neighborhoods under ideal simulation; under noise this structure compresses toward the maximally mixed state while preserving its relative ordering.

These results demonstrate a scalable, reproducible workflow for quantum kernel experimentation at HEP-relevant scale, quantifying the practical cost of realistic hardware noise on quantum-enhanced classification.
\end{abstract}




\end{frontmatter}





\section{Introduction}

Accurate classification of particle interactions is central to many areas of high-energy physics (HEP), including signal–background discrimination, searches for rare processes, and real-time triggering. Modern detectors such as those used in the CLAS12 experiment generate large volumes of structured yet highly correlated data, motivating the adoption of machine learning (ML) techniques for event selection and feature extraction. Classical approaches—including support vector machines (SVMs), boosted decision trees, and ensemble methods—have become standard tools due to their robustness and interpretability. However, many physics datasets contain nonlinear or entangled correlations that can be difficult for classical models to capture, particularly when the features are compact, highly structured, or derived from geometric detector responses.

Quantum machine learning (QML) offers a potential alternative by exploiting the exponentially large state space of quantum systems to embed and compare classical data. Quantum kernel methods, and the quantum support vector machine (QSVM) in particular, construct a kernel through a parameterized quantum circuit that maps input features into a high-dimensional Hilbert space. This enables the evaluation of similarities between data points via quantum state fidelities, providing a quantum analogue to the classical “kernel trick.” Theoretical work suggests that certain quantum kernels may be difficult to reproduce classically, but the degree to which these advantages manifest in practical scientific tasks — and the degree to which they survive the noise present on real quantum hardware — remains an open question.

In this work, we investigate the use of a QSVM for particle–track classification using wire-chamber data from the CLAS12 detector at Jefferson Lab, under both ideal (noiseless) simulation and a calibrated hardware noise model. Each event is compressed into a six-dimensional vector of intensity-weighted centroids across segmented drift-chamber regions, preserving key geometric information about the track while enabling efficient encoding. These features are mapped to six-qubit quantum states through a fully entangling ZZFeatureMap, and the corresponding quantum kernel is computed via statevector simulation using Qiskit Aer for the ideal case, and via a compute-uncompute measurement protocol under a FakeMumbaiV2 noise model for the noisy case.

We benchmark the QSVM against four classical SVM kernels (linear, polynomial, RBF, sigmoid) as well as extremely randomized trees (ERT), the current state-of-the-art for CLAS12 track identification. Under ideal simulation, the QSVC achieves the highest recall of any model evaluated, with only a handful of false negatives among more than $4\times 10^5$ test events, while remaining within one to two percentage points of the best classical model on every other metric. This extremely low false-negative rate is a highly desirable property in HEP, where missed signal events directly reduce sensitivity to rare or exotic physics. Classical models can match the QSVC's accuracy only by sacrificing recall, indicating that they do not capture the same nonlinear correlations present in the detector-derived feature space. Under a realistic hardware noise model, this recall advantage is attenuated but persists: the noisy QSVC retains a recall of 99.51\%, exceeding every classical baseline evaluated here except the RBF kernel, even as overall accuracy and background rejection degrade substantially.

Beyond performance metrics, we analyze the geometry of the quantum embedding through statevector statistics, entanglement entropy, fidelity distributions, class-resolved kernel fidelity distributions, and dimensionality-reduction visualizations. These analyses reveal that the ZZFeatureMap produces a structured, expressive embedding in which inter-class states become nearly orthogonal while intra-class states form coherent neighborhoods— precisely the geometry that enables the QSVC to form a sharp, high-margin decision boundary under ideal simulation. Extending this analysis to the noisy regime shows that hardware noise compresses this class-dependent structure toward the maximally mixed state without erasing its relative ordering, providing a geometric explanation for why classification performance degrades gracefully rather than collapsing to chance.

Taken together, these results demonstrate that quantum-enhanced kernels can extract complex correlations from compact detector features and provide a practical performance advantage even when executed on classical simulators, while also establishing a quantitative bound on how much of that advantage survives under realistic circuit-level noise. This work establishes QSVCs as a promising tool for high-fidelity event classification in HEP and motivates future exploration of hardware-efficient encodings, error-mitigated kernel evaluation, hybrid quantum–classical pipelines, and applications to real-time or streaming environments.

The main contributions of this work can be summarized as follows:

\begin{enumerate}

\item \textbf{Application of quantum kernel methods to CLAS12 track classification.}
To our knowledge, we present the first study applying a quantum support vector classifier (QSVC) to centroid-based
features derived from the CLAS12 drift-chamber detector, demonstrating the feasibility of
quantum kernel methods for realist particle--track classification.

\item \textbf{Quantum embedding and scalable, multi-node kernel-evaluation workflow.}
A six-dimensional detector representation is encoded into a six-qubit quantum state using a
fully entangling \texttt{ZZFeatureMap}, with quantum-state fidelities used as the kernel
similarity measure. By decoupling quantum-state preparation from kernel construction through
cached statevector embeddings and block-wise evaluation distributed across a multi-node
MPI-based HPC allocation, the approach scales to datasets
containing more than $5\times10^{5}$ events.

\item \textbf{Systematic comparison with classical baselines and signal retention advantage.}
We benchmark the QSVC against support vector machines with linear, polynomial, RBF, and
sigmoid kernels, as well as extremely randomized trees (ERT). The QSVC achieves near-perfect
recall while ERT achieves higher precision and specificity, consistent with smoother
similarity-based decision boundaries that better capture structured geometric correlations at
the cost of some background contamination. This advantage persists, in attenuated form, under
realistic hardware noise.

\item \textbf{Direct characterization of performance under hardware noise.}
Using a calibrated hardware noise model (FakeMumbaiV2) and a compute / uncompute measurement
protocol, we evaluate a noisy quantum kernel on the same dataset and quantify the resulting
degradation in AUC, significance improvement, and background rejection relative to the ideal
simulation, while showing that the QSVC's signal-retention behavior is attenuated rather than
eliminated by circuit-level decoherence.

\item \textbf{Geometry-driven analysis of the induced quantum feature space, ideal and noisy.}
We analyze the structure of the quantum embedding using entanglement entropy, fidelity
distributions, class-resolved kernel fidelity distributions, and dimensionality-reduction visualizations, showing how
the feature map produces coherent intra-class neighborhoods and strong inter-class separation
under ideal simulation, and how this structure is compressed — but not erased — under hardware noise.

\end{enumerate}

Read together, these contributions form a single narrative rather than five independent results: the scalable HPC workflow makes it possible to evaluate the quantum kernel at a scale where its classification behaviour can be measured with statistical confidence, the geometric analysis explains why that behaviour takes the specific form it does — near-complete signal retention driven by class-dependent fidelity structure — and the noise characterization shows that this same geometric mechanism degrades gracefully rather than collapsing under realistic hardware decoherence. The result is not simply a demonstration that a QSVC can be applied to CLAS12 data, but a mechanistic account of when quantum kernel methods offer a distinct advantage over classical baselines and how much of that advantage a real device is likely to preserve. The multi-node HPC workflow underlying this scale is consistent with our prior work on distributing NISQ (D-NISQ) quantum simulation across classical HPC infrastructure to reach utility-scale problem sizes~\cite{billias2026hybridquantumclassicalframeworkutilityscale}, applied here to sample-level rather than circuit-level decomposition.

\subsection*{Regimes of Advantage for Quantum Kernel Methods}

The results of this study, together with recent findings in the literature, suggest that
quantum kernel methods provide the greatest benefit in the following regimes:

\begin{itemize}
\item \textbf{Low–false-negative operating regions:} where signal retention is critical
for maintaining sensitivity in rare-event searches, a property we find to be only partially
degraded under realistic hardware noise.

\item \textbf{Structured geometric feature spaces:} where detector-derived features encode
spatial or angular correlations that can be naturally represented through entangling feature maps.

\item \textbf{Data-limited scenarios:} where quantum embeddings may achieve strong
performance with reduced training data through more expressive similarity representations
— a hypothesis motivated by results reported elsewhere in the literature, which this
study's fixed 100,488-sample ideal training set was not designed to test directly.

\item \textbf{High-recall requirements:} where preserving signal events is essential and
decision-boundary stability is critical.
\end{itemize}

\section{Related Work}
Efforts to modernize particle-track reconstruction in the CLAS12 detector at Jefferson Lab have increasingly incorporated machine learning (ML) techniques, reflecting a broader transition from traditional reconstruction algorithms toward AI-assisted analysis workflows. Early applications focused on reducing the combinatorial complexity of drift-chamber tracking. Thomadakis et al.~\cite{Thomadakis2022} demonstrated that classical machine-learning models, including convolutional neural networks (CNNs), multilayer perceptrons, and extremely randomized trees (ERTs), could act as intelligent pre-selectors for track candidates, substantially reducing reconstruction time while maintaining physics performance. Subsequent studies extended ML beyond offline classification into high-throughput environments. Gavalian~\cite{gavalian2024realtimechargedtrackreconstruction} showed that kinematic quantities such as momentum and scattering angles could be inferred directly from raw cluster information, enabling near real-time event characterization suitable for trigger-level applications. Additional efforts addressed detector noise through CNN-based denoising autoencoders~\cite{Thomadakis2022_denoising}, developed ML-driven Level-3 trigger systems for electron identification~\cite{gavalian2022clas12trackreconstructionartificial}, and expanded reconstruction pipelines toward full event-topology classification and reaction identification~\cite{Thomadakis2023_CLAS12ML}. Collectively, these studies established machine learning as an integral component of modern CLAS12 reconstruction and analysis workflows.

In parallel, growing interest in quantum machine learning (QML) has motivated investigations into whether quantum feature spaces can provide advantages for classification tasks in high-energy physics (HEP). Early proof-of-principle studies demonstrated that quantum support vector machines (QSVMs), variational quantum classifiers (VQCs), and quantum kernel estimation techniques could be applied to realistic collider analyses using reduced kinematic feature sets encoded into low-qubit quantum circuits~\cite{Havlicek2019,Chan2020,Wu2021}. A major milestone was the application of QSVM and VQC models to Higgs boson analyses at the LHC using IBM quantum simulators and superconducting quantum hardware~\cite{Chan2020,Wu2021}. These studies investigated processes including $t\bar{t}H$ production and $H\rightarrow\mu^+\mu^-$ decay using PCA-compressed kinematic observables encoded into quantum feature spaces, demonstrating classification performance competitive with classical support vector machines and boosted decision trees while establishing the practical feasibility of quantum-kernel workflows on NISQ-era hardware. Notably, Wu et al.~\cite{Wu2021} scaled their $t\bar{t}H$ QSVM-Kernel study up to 50,000 events using 15--20 qubits across Google, IBM, and Amazon quantum simulation frameworks — to our knowledge the largest dataset used in a published HEP quantum-kernel study prior to this work — and found the QSVM-Kernel method to perform comparably to, rather than exceeding, its classical SVM and BDT counterparts at that scale. The present work extends this line of investigation by more than an order of magnitude in dataset size (Section~\ref{sec:evaluation}), enabled by the multi-node HPC architecture described in Section~\ref{sec:hpc}, and identifies a distinguishing quantum-kernel advantage in signal-retention behaviour rather than in raw ranking metrics such as AUC, where classical baselines remain competitive.

Since these initial demonstrations, QML research in HEP has expanded substantially. Quantum kernel methods have been investigated for resonance detection at BESIII~\cite{BESIII_QSVM}, continuum suppression at Belle II~\cite{Heredge2021}, Higgs classification, anomaly detection, graph-based tracking architectures, and hybrid quantum--classical learning frameworks~\cite{Schuhmacher2023,Cappelli2024,Guan2021}. In particular, quantum graph neural network approaches have been explored for charged-particle tracking and detector-geometry-aware learning tasks, reflecting a broader shift from low-dimensional event-level observables toward structured detector representations. Across these applications, quantum feature maps consistently produce structured Hilbert-space embeddings in which inter-class states become increasingly separated while intra-class states retain coherent similarity structure~\cite{Schnabel_2025}. Several studies report competitive classification performance using reduced training datasets~\cite{BESIII_QSVM}, improved background rejection at fixed signal efficiency~\cite{Heredge2021}, and enhanced sensitivity to physically meaningful correlations encoded in detector observables~\cite{okawa2025quantumartificialintelligencepattern}.

Large collaborative programs at Brookhaven National Laboratory (BNL), CERN, Fermilab, and DESY have further advanced the study of QML for HEP applications, investigating both quantum kernel methods and variational quantum architectures for Higgs boson identification, jet tagging, and track reconstruction~\cite{BNL_QML,CERN_QML,DESY_QML}. Across these efforts, QSVM-based approaches have generally achieved performance comparable to established classical methods such as boosted decision trees, particularly in analyses involving rare processes and high-dimensional correlations. In contrast, variational quantum classifiers and quantum neural networks frequently exhibit sensitivity to circuit design choices and training instabilities, highlighting the relative robustness of kernel-based approaches within current NISQ limitations. A recurring conclusion across these studies is that performance is strongly linked to the geometry induced by quantum embeddings, which leverage high-dimensional Hilbert spaces to construct expressive similarity measures while naturally accommodating structured detector information~\cite{QML_HEP_Review}.

These observations align with broader comparisons between kernel-based and tree-based learning methods in HEP~\cite{Choudhury_2024,Hofmann_2008}. Tree-based ensemble models partition feature spaces through recursive axis-aligned decision boundaries and often achieve excellent overall classification accuracy. However, kernel methods instead perform global margin maximization in an implicit feature space, yielding smoother decision boundaries that can better capture nonlinear physical manifolds present in detector data. Prior studies have associated these geometric properties with improved robustness to noise, enhanced generalization in structured feature spaces, and stronger background suppression at fixed signal efficiency~\cite{6177282,10069444}. Such characteristics are particularly attractive for rare-event searches and high-recall classification tasks, where preserving signal efficiency is critical.

Despite these advances, most HEP-QML demonstrations remain constrained by limited qubit counts, shallow circuit depths, and the computational expense of quantum kernel evaluation, and relatively few directly quantify how classification performance degrades under realistic hardware noise on the same dataset used for ideal-simulation benchmarking. Consequently, many existing studies rely on relatively small datasets and heavily compressed feature representations. The present work extends these investigations to a large-scale CLAS12 drift-chamber dataset containing more than $5\times10^5$ events. Using detector-derived centroid features encoded through a fully entangling \texttt{ZZFeatureMap}, we evaluate whether quantum kernel methods can effectively capture geometric correlations present in detector occupancy patterns, and additionally provide a direct, same-dataset comparison between ideal-simulation performance and performance under a calibrated hardware noise model (FakeMumbaiV2), quantifying the gap that future hardware-execution and error-mitigation studies would need to close. This provides a direct connection between the established CLAS12 machine-learning program and the emerging field of quantum-enhanced classification for high-energy physics.

\section{Support Vector Machines}

Support Vector Machines (SVMs) are supervised learning models for binary classification. 
Given labeled data $\{(x_i, y_i)\}_{i=1}^N$ with $x_i \in \mathbb{R}^d$ and 
$y_i \in \{-1,+1\}$, an SVM seeks a decision function of the form

\begin{equation}
    f(x) = w^T \phi(x) + b
\end{equation}

where $\phi(x)$ denotes a feature mapping and 
$w$ and $b$ define a separating hyperplane in the feature space.

The model is obtained by solving the soft margin optimization problem

\begin{equation}
    \min_{w,b,\xi} \frac{1}{2}\|w\|^2 + C \sum_{i=1}^N \xi_i
\end{equation}

subject to

\begin{equation}
    y_i f(x_i) \ge 1 - \xi_i, \quad \xi_i \ge 0
\end{equation}

where $C > 0$ controls the trade-off between margin maximization and 
classification error.

In its dual formulation, the solution depends only on inner products 
between feature vectors, enabling the use of kernel functions 
$K(x_i, x_j) = \phi(x_i)^T \phi(x_j)$. 
The resulting classifier can be written as

\begin{equation}
    f(x) = \sum_{i \in SV} \alpha_i y_i K(x_i, x) + b
\end{equation}

where the sum runs over the support vectors.

When applied to quantum-processed data, the feature map $\phi(x)$ may correspond to a quantum state preparation or circuit-based embedding, with kernel values obtained from the squared inner product between quantum states encoding the data. In this setting, the SVM operates in a feature space defined implicitly by quantum transformations, enabling classification based on quantum-enhanced representations. Under ideal, noiseless simulation, $K(x_i,x_j)$ can be evaluated exactly from cached statevectors; under realistic hardware conditions, each kernel entry must instead be estimated from repeated circuit measurement, introducing both statistical shot noise and systematic error from gate infidelity and decoherence. Section~\ref{sec:noisy_sim} details how this distinction is implemented, and Sections~\ref{sec:evaluation}--\ref{sec:data_analysis} quantify its effect on classification performance.

While ensemble methods such as random forests or extremely randomized trees often achieve strong classification accuracy, support vector machines offer several properties that make them particularly well suited for kernel-based approaches. SVMs construct decision boundaries by maximizing the margin between classes in a transformed feature space defined implicitly through a kernel function. This formulation allows nonlinear correlations between features to be captured through pairwise similarities between samples rather than through large ensembles of axis aligned decision rules.

Because detector observables vary continuously with particle trajectories and detector geometry, classification models with smooth decision boundaries are often better aligned with the underlying data structure. Kernel-based similarity methods naturally produce such smooth boundaries, whereas tree based methods partition the feature space into piecewise constant regions. Moreover, the kernel formulation makes SVMs a natural framework for incorporating quantum feature maps: replacing the classical kernel with a fidelity based quantum kernel preserves the margin maximization principle while allowing the feature space itself to be defined by a quantum embedding. We note that this theoretical distinction motivates the comparison in this work, but does not by itself predict which approach will perform better empirically; Section~\ref{sec:evaluation} shows that the two paradigms in fact produce complementary rather than strictly ordered results on this dataset, with the tree-based ERT model achieving the strongest overall ranking performance.

\section{Methods}

We evaluate a quantum support vector classifier (QSVC) for particle–track
discrimination using quantum kernels constructed from a fully entangling
ZZFeatureMap, under both ideal (noiseless) simulation and a realistic
hardware noise model. The workflow consists of five main stages: data
preprocessing, quantum feature encoding, statevector simulation, noisy
kernel evaluation via a compute-uncompute measurement protocol, and
kernel-based classification. All experiments were executed on a
five-node high-performance computing allocation (one MPI rank per node,
40 CPU cores per node, 200 cores total) managed via SLURM, with
statevector generation and ideal kernel matrix construction distributed
across ranks and exploiting NumPy's BLAS-accelerated linear algebra
routines within each rank. Section~\ref{sec:hpc} details this multi-node
execution architecture and the scaling behaviour observed across pipeline
runs.

\begin{figure}
    \centering
    \includegraphics[width=0.75\linewidth]{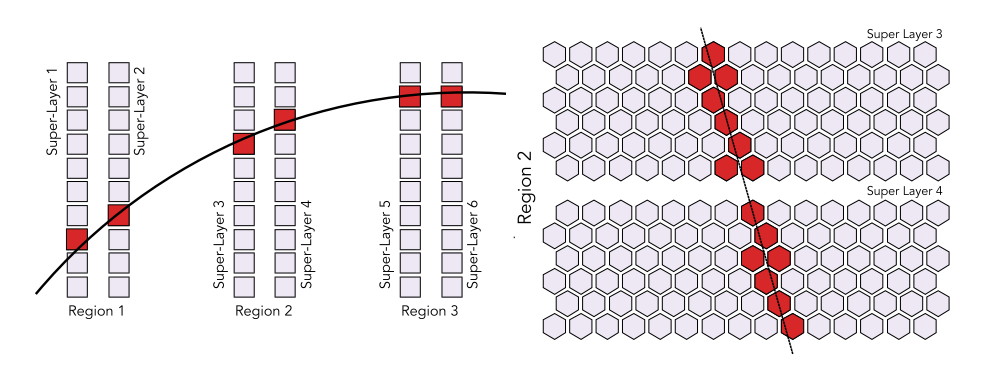}
    \caption{Wire activations in a drift-chamber superlayer produced by a
    traversing charged particle. Activated segments (left) and individual hit
    cells (right) illustrate the spatial structure from which centroid-based
    features are derived \cite{gavalian2022clas12trackreconstructionartificial}.}
    \label{fig:chamber}
\end{figure}

\subsection{Data Preprocessing}

The raw input consists of \(6 \times 112\)-pixel drift-chamber
``superclusters,'' each corresponding to one of the six longitudinal
superlayers of the CLAS12 drift chamber (Fig.~\ref{fig:chamber}). To obtain
a compact yet informative representation, we compute an intensity-weighted
centroid within each superlayer. For a superlayer with pixel intensities
\(I_k\) and wire indices \(w_k\), the centroid is

\begin{equation}
    c = \frac{\sum_k I_k\, w_k}{\sum_k I_k},
\end{equation}

normalized to the interval \([0,1]\). Each event is thus represented by a
six-dimensional feature vector that preserves the global geometry of the
underlying track while substantially reducing input dimensionality. We
acknowledge that this compression discards higher-order spatial information —
including wire-level hit multiplicity and per-superlayer occupancy patterns —
that may carry additional discriminating power. The centroid representation
was chosen to match the compact feature encoding used in prior CLAS12 machine
learning studies~\cite{Thomadakis2022}, enabling direct comparison of
the quantum kernel approach within an established feature space. Investigating
richer representations, including full supercluster images or graph-based
detector encodings, is reserved for future work.

The dataset used in this study consists exclusively of negatively charged
particles, reflecting a data-availability constraint at the time of this
analysis. This single-charge restriction simplifies the physical manifold
relative to mixed-charge CLAS12 tracking workflows, as all tracks curve in
the same direction within the toroidal magnetic field, reducing the geometric
variance that classifiers must resolve. Consequently, performance metrics
reported here should not be interpreted as directly comparable to published
mixed-charge baselines. All baseline classifiers evaluated in this work —
including ERT — were trained and tested on the same
single-charge dataset to ensure a fair comparison. Future work will
incorporate mixed-charge datasets within the same evaluation framework as
such samples become available.

\subsection{Quantum Feature Encoding}

Each feature vector \(x \in \mathbb{R}^6\) is embedded into a six-qubit
quantum state via the ZZFeatureMap with \texttt{reps=1} and full
entanglement, applying data-parameterized single-qubit rotations followed by
pairwise ZZ entangling gates across all qubit pairs. The encoded state is

\begin{equation}
    |\psi(x)\rangle = U_{\mathrm{ZZF}}(x)\,|0\rangle^{\otimes 6},
\end{equation}

where \(U_{\mathrm{ZZF}}\) denotes the feature-map circuit. This fully
entangling encoding introduces nonlinear correlations among features and
maps each event into a \(2^6 = 64\)-dimensional complex Hilbert space.

We note that for \(n = 6\) qubits the statevectors are 64-dimensional
complex vectors, and the ideal quantum kernel
\(K(x_i, x_j) = |\langle\psi(x_i)|\psi(x_j)\rangle|^2\) is analytically
equivalent to the classical inner product \(|XX^\dagger|^2\) once
statevectors are cached. At this qubit count, the ideal kernel can therefore
be evaluated classically without quantum hardware. The ZZFeatureMap circuit and
Qiskit Aer simulation backend are used not because classical evaluation is
intractable, but because this workflow is directly portable to larger qubit
registers, alternative feature maps, and realistic device noise models
without any modification to the kernel construction or classification stages.
This design choice prioritises generalizability and reproducibility over
raw execution speed at the present scale. The noisy kernel evaluation
described in Section~\ref{sec:noisy_sim} does not share this
classical-equivalence shortcut, since it requires explicit circuit
execution and measurement rather than a cached inner product.

\subsection{Statevector Simulation}

Encoded quantum states are evaluated once using Qiskit Aer's ideal
statevector simulator via \texttt{Statevector.from\_instruction()}.
Statevectors are computed in parallel and stored in compressed \texttt{.npz}
format, decoupling the circuit simulation step from all downstream kernel
construction and classification. This caching strategy means that
hyperparameter sweeps, alternative SVM configurations, and all geometric
analyses reported in Section~\ref{sec:data_analysis} reuse the same precomputed
embeddings without rerunning circuit simulation, making large-scale
experimentation tractable.

The ideal statevector simulation described above does not incorporate
device noise, finite coherence times, or gate error rates characteristic
of current NISQ hardware, and establishes an upper bound on achievable
performance under the ZZFeatureMap encoding. To characterise performance
degradation under realistic hardware conditions, we additionally evaluate
a noisy quantum kernel using Qiskit's
\texttt{NoiseModel.from\_backend()} interface, described in
Section~\ref{sec:noisy_sim}. Results from both the ideal and noisy
studies are reported and compared directly throughout
Sections~\ref{sec:evaluation}--\ref{sec:data_analysis}.

\subsection{Noisy Simulation via Hardware Error Modeling}
\label{sec:noisy_sim}

Unlike the ideal statevector kernel, which is evaluated as a cached inner
product between statevectors, the noisy kernel requires explicit circuit
execution: each entry \(K_{ij}\) is estimated from repeated measurement of
a compute-uncompute circuit \(U_{\mathrm{ZZF}}(x_j)^\dagger
U_{\mathrm{ZZF}}(x_i)\,|0\rangle^{\otimes 6}\), with \(K_{ij}\) taken as
the observed frequency of the all-zeros outcome over 1{,}024 shots. A
noise model was constructed via \texttt{NoiseModel.from\_backend()} using
the FakeMumbaiV2 fake-backend calibration snapshot, which reproduces
representative single- and two-qubit gate error rates, readout error, and
decoherence characteristics of a real superconducting device. The
compute-uncompute circuit template was transpiled at
\texttt{optimization\_level=3} to a fixed physical layout, yielding a
circuit of depth 58 (58 CX, 52 RZ, 12 SX, 6 measurement, and 1 barrier
operation); all subsequent per-pair circuits were transpiled at
\texttt{optimization\_level=1} against this fixed layout to avoid
re-optimizing an already-fixed topology.

Because circuit-level noisy simulation is substantially more expensive
than the cached-statevector ideal kernel, the noisy study uses a reduced
subsample of 500 training and 2{,}000 test events drawn from the same
preprocessed dataset. The training kernel \(K_{\mathrm{train}}\) is
symmetric by construction, so only the 500 diagonal (self-fidelity) and
124{,}750 upper-triangular off-diagonal entries were evaluated explicitly
— 125{,}250 circuit evaluations rather than the full \(500^2 = 250{,}000\)
— with the lower triangle populated by symmetry. Diagonal entries were
computed explicitly rather than assumed equal to 1, since decoherence
causes even a state's self-overlap to fall measurably below unity; the
resulting noisy kernel has mean diagonal value substantially less than 1
and overall mean entry \(\approx 0.11\), consistent with noise-driven
concentration of the encoded states toward the maximally mixed state
(Section~\ref{sec:data_analysis}). The rectangular test kernel
\(K_{\mathrm{test}} \in \mathbb{R}^{2{,}000 \times 500}\) admits no such
symmetry and required the full \(1{,}000{,}000\) circuit evaluations.
Circuit dispatch within each MPI rank was parallelized across a pool of
40 worker processes, with circuits transpiled and submitted in fixed-size
chunks to bound peak memory. Section~\ref{sec:hpc} describes how this
workload was further distributed across MPI ranks for the test-kernel
phase.

The resulting noisy kernel matrices, along with the noisy QSVC trained and
evaluated on them, are compared against the ideal results throughout this
work (Sections~\ref{sec:evaluation}--\ref{sec:data_analysis}), providing a
direct, quantitative characterization of how circuit-level decoherence at
current NISQ-era noise levels degrades quantum-kernel classification
performance relative to the ideal-simulation upper bound.

\subsection{Quantum Kernel Construction}

Classification is performed using an SVM with a quantum kernel defined as
the squared fidelity between encoded states:

\begin{equation}
    K(x_i, x_j) = |\langle \psi(x_i) \mid \psi(x_j) \rangle|^2.
\end{equation}

Let \(X_{\mathrm{train}} \in \mathbb{C}^{N_{\mathrm{train}} \times 64}\)
and \(X_{\mathrm{test}} \in \mathbb{C}^{N_{\mathrm{test}} \times 64}\)
denote the matrices of cached training and test statevectors respectively.
The ideal kernel matrices are

\begin{equation}
    K_{\mathrm{train}} = |X_{\mathrm{train}} X_{\mathrm{train}}^\dagger|^2,
    \qquad
    K_{\mathrm{test}} = |X_{\mathrm{test}} X_{\mathrm{train}}^\dagger|^2,
\end{equation}

where \(K_{\mathrm{test}} \in \mathbb{R}^{n_{\mathrm{test}} \times
n_{\mathrm{train}}}\) provides the pairwise similarity between each test
event and all training events. Both matrix products are evaluated as single
BLAS-accelerated \texttt{numpy} operations.

For \(N_{\mathrm{train}} = 100{,}488\) samples,
\(K_{\mathrm{train}}\) is a \(100{,}488 \times 100{,}488\)
double-precision matrix requiring approximately 80\,GB of RAM. Each of
the 5 MPI ranks computes its own row-block of \(K_{\mathrm{train}}\)
locally before rank 0 gathers all blocks into the full matrix. This
gather step dominates the training phase: the full
\texttt{ideal\_svm\_training} phase required 647\,s (10.8\,min) of
wall-clock time, of which approximately 617\,s (95\%) was consumed by
MPI-gathering the four remaining row-blocks from ranks 1--4 onto rank 0,
and only the final \(\sim\)30\,s corresponded to the LIBSVM
quadratic-programming solve itself once the full precomputed kernel was
available. The assembled kernel was passed directly to Scikit-Learn's
\texttt{SVC} with \texttt{kernel=\textquotedbl precomputed\textquotedbl},
which internally uses the LIBSVM solver to solve the standard SVM dual
quadratic programme over the full kernel. No kernel approximation (e.g.\
Nystr\"om) was employed. The SVM was trained with the default
regularisation parameter \(C = 1.0\). We acknowledge that the
\(\mathcal{O}(N^2)\) memory footprint and \(\mathcal{O}(N^2)\)–%
\(\mathcal{O}(N^3)\) solver complexity represent the primary scalability
limitation of the present approach, and we discuss mitigation strategies —
including approximate kernel methods and sparse SVM solvers — in
Section~\ref{sec:discussion}.

\subsection{Classical Baselines}

To situate the QSVC within the broader CLAS12 machine learning landscape,
we evaluate it against five classical baselines, all trained and tested on
the same single-charge dataset and train/test split.

\textbf{Classical SVM kernels.} Four standard SVM kernels are evaluated
using Scikit-Learn's \texttt{SVC}: linear, polynomial, RBF, and sigmoid.
These provide a direct comparison of the quantum kernel against established
classical kernel functions within the same SVM framework and on the same
feature representation.

\textbf{Extremely Randomized Trees (ERT).} An \texttt{ExtraTreesClassifier}
with 300 estimators, entropy criterion, and all features considered at each
split is trained on the raw six-dimensional centroid features. ERT represents
the current state-of-the-art for CLAS12 track identification in the
literature~\cite{Thomadakis2022}.

Notably, these classical baselines — none of which involve any quantum
simulation — dominate the total wall-clock cost of the pipeline. The four
classical SVM kernels and ERT together required 13{,}466\,s (3.74\,h,
69\% of total pipeline time), more than ten times longer than the entire
ideal QSVC path (statevector generation through test inference, 1{,}289\,s
combined). Within the classical baseline suite, the sigmoid kernel was the
single slowest component (82.2\,min), followed by the polynomial kernel
(64.6\,min), linear kernel (50.5\,min), and RBF kernel (27.1\,min); ERT
itself trained in under 5\,s. This asymmetry — an expressive quantum
kernel evaluated in minutes once statevectors are cached, versus
general-purpose classical SVM kernels requiring hours on the same
100{,}488-sample training set — is a direct consequence of the
statevector-caching strategy described in Section~\ref{sec:noisy_sim} and
is discussed further in Section~\ref{sec:discussion}.

\subsection{Chunked Evaluation and Resource Constraints}

Because the test kernel \(K_{\mathrm{test}}\) is large but does not need
to reside in memory simultaneously, ideal test-time inference uses a
chunked evaluation strategy. The test set is processed in blocks of
5{,}000 samples:

\begin{enumerate}
    \item compute a submatrix of \(K_{\mathrm{test}}\) for the current block,
    \item evaluate SVM decision scores and predicted labels for that block,
    \item accumulate confusion-matrix contributions across all blocks.
\end{enumerate}

This strategy makes QSVC inference over \(4 \times 10^5\) test events
feasible within standard memory limits while producing results mathematically
identical to full-matrix evaluation. We emphasise that chunking applies only
to test inference; training requires the full \(K_{\mathrm{train}}\) matrix
to be resident in memory simultaneously, as described above.

\subsection{Multi-Node HPC Execution and Scaling}
\label{sec:hpc}

The multi-node architecture described in this section follows the same
underlying principle as our prior work on D-NISQ quantum
simulation across classical HPC infrastructure to reach utility-scale
problem sizes~\cite{billias2026hybridquantumclassicalframeworkutilityscale},
which decomposes a single quantum circuit both at the data level (into
sub-images) and the circuit level (into sub-circuits) to make large-scale
NISQ simulation tractable on a distributed classical cluster. The present
work applies the same distributed-simulation philosophy along a different
axis: rather than decomposing individual circuits, we decompose the
\emph{sample set} — statevector generation and kernel-matrix construction
are partitioned into row-blocks distributed across MPI ranks, with a
further layer of intra-rank parallelism for noisy circuit dispatch, as
described below.

The full pipeline — ideal statevector generation, ideal kernel
construction and classification, all five classical baselines, and the
noisy kernel study — was executed as a single SLURM job spanning 5 nodes
(one MPI rank per node, 40 CPU cores per node, 200 cores total). Two
levels of parallelism are used concurrently: MPI distributes row-blocks
of work across ranks/nodes, while within each rank a pool of 40 worker
processes handles batched, per-pair circuit transpilation and dispatch
for the noisy kernel phase. Ideal statevector generation is similarly
distributed, with each rank generating and locally caching its own
\(\sim\)100,488-row block of the full 502,443-event statevector array
before rank 0 assembles the complete array for downstream use.

The noisy test kernel \(K_{\mathrm{test}} \in
\mathbb{R}^{2{,}000\times 500}\) — comprising 1{,}000{,}000 circuit
evaluations at 1{,}024 shots each — was distributed across all 5 MPI
ranks, with each rank computing 400 of the 2{,}000 test rows against the
full 500-sample training set. Rank 0's share (400 rows,
200{,}000 pairs) completed in 3{,}008\,s. Combined with the symmetric
125{,}250-pair training kernel (Section~\ref{sec:noisy_sim}), the full
noisy kernel study — construction, SVM training, and inference — added
4{,}779\,s (79.7\,min) to the pipeline. Without this two-level
parallelization (MPI across ranks for row-block distribution, plus
per-rank worker pools for circuit dispatch), evaluating even this
reduced-scale 500/2{,}000 noisy study would not have been tractable
within a comparable wall-clock budget.

Total wall-clock time for the complete pipeline — ideal statevector
generation, ideal QSVC training and inference, all classical baselines,
and the full noisy kernel study (construction, training, and inference)
— was 19{,}589.70\,s (5.44\,h). Of this, the noisy simulation phases
contributed 4{,}779\,s (79.7\,min, 24.4\% of total), the ideal QSVC path
contributed 1{,}289\,s (6.6\%), and the classical baseline suite
contributed the remaining 13{,}466\,s (68.8\%), as noted above.

Together, these methods enable the application of quantum kernels to
realistic detector data at scale, balancing the expressive power of
quantum feature maps with the computational constraints of both classical
simulation and noisy circuit-level evaluation.

\section{Evaluation}
\label{sec:evaluation}

We evaluated the QSVC model on a dataset of 502,443 events using a 20\%
train and 80\% test split with a fixed random seed (seed~=~49). After
preprocessing and encoding the data with a fully entangled ZZFeatureMap
(\texttt{reps=1}, full entanglement), the model was trained on 100,488
samples and tested on 401,955 unseen examples. All classical baselines
were trained and evaluated on the same split to ensure a fair comparison.
A separate noisy-simulation study, using the FakeMumbaiV2 hardware noise
model described in Section~\ref{sec:noisy_sim}, was trained on 500 samples
and tested on 2,000 samples drawn from the same preprocessed dataset; the
substantially smaller sample size reflects the computational cost of
circuit-level noisy kernel evaluation relative to the cached-statevector
ideal kernel, as discussed in Section~\ref{sec:hpc}.

\subsection{Ideal QSVC Performance}
\label{sec:ideal_eval}

The training confusion matrix reveals an extremely low false-negative rate,
with only 8 missed signal events out of over 100,000 training examples,
indicating that the model reliably identifies genuine signal tracks during
fitting:

\begin{table}[htbp]
\centering
\begin{tabular}{lcc}
\toprule
\textbf{Predicted / Actual} & Background & Signal \\
\midrule
Background & 48006 & 8 \\
Signal     & 2177  & 50297 \\
\bottomrule
\end{tabular}
\caption{Training confusion matrix for the ideal QSVC. Rows indicate predicted
class; columns indicate true class.}
\label{tab:training_confusion}
\end{table}

On the test set, the ideal QSVC maintains this behaviour, producing only 13
false negatives across 401,955 events. This yields a near-perfect signal
recall of \(\approx 99.99\%\), underscoring the model's capacity to retain
essentially all genuine signal events. The corresponding false-positive
count of 9,053 gives a precision of \(\approx 95.7\%\) and a specificity
of \(\approx 95.5\%\):

\begin{table}[htbp]
\centering
\begin{tabular}{lcc}
\toprule
\textbf{Predicted / Actual} & Background & Signal \\
\midrule
Background & 191985 & 13 \\
Signal     & 9053   & 200904 \\
\bottomrule
\end{tabular}
\caption{Test confusion matrix for the ideal QSVC. Rows indicate predicted class;
columns indicate true class.}
\label{tab:test_confusion}
\end{table}

Overall, only 21 signal events are misclassified across training and test
combined, reflecting a strong and consistent bias toward signal retention —
a deliberately favourable property in high-energy physics, where missed
signal events directly reduce sensitivity to rare or exotic processes and
cannot be recovered downstream.

\subsection{Noisy QSVC Performance}
\label{sec:noisy_eval}

Under the FakeMumbaiV2 noise model, the QSVC was trained on 500 events and
evaluated on a held-out test set of 2,000 events:

\begin{table}[htbp]
\centering
\begin{tabular}{lcc}
\toprule
\textbf{Predicted / Actual} & Background & Signal \\
\midrule
Background & 782 & 5 \\
Signal     & 201 & 1012 \\
\bottomrule
\end{tabular}
\caption{Test confusion matrix for the noisy QSVC (500 train / 2,000 test,
FakeMumbaiV2 noise model). Rows indicate predicted class; columns indicate
true class.}
\label{tab:noisy_confusion}
\end{table}

The noisy QSVC achieves an accuracy of 0.8970, precision of 0.8343, recall
of 0.9951, specificity of 0.7955, and F1 score of 0.9077. Strikingly, the
model's signal-retention bias survives circuit-level decoherence: recall
falls only marginally, from 99.99\% under ideal simulation to 99.51\% under
noise (5 missed signal events out of 1,017), even as overall accuracy drops
by roughly 8 percentage points and specificity falls by nearly 16 points.
This indicates that hardware noise degrades the QSVC's ability to reject
background substantially more than its ability to retain signal — consistent
with the compressed but ordering-preserving kernel fidelity structure
examined in Section~\ref{sec:data_analysis}. We caution that the noisy
result is evaluated on a much smaller test set (2,000 events, versus
401,955 for the ideal and classical models); the confusion-matrix counts
in Table~\ref{tab:noisy_confusion} should therefore be read with
correspondingly wider statistical uncertainty than the ideal results above.

\subsection{Comparison with Classical Baselines}

Table~\ref{tab:classification_results} extends this evaluation by comparing
the QSVC to four classical SVM kernels (linear, polynomial, RBF, sigmoid)
and extremely randomized trees (ERT), the current state-of-the-art
classifier for CLAS12 track identification~\cite{Thomadakis2022}, together
with the noisy QSVC introduced above.
All classical models and the ideal QSVC were trained on the same negatively
charged single-charge dataset and evaluated on the same held-out
401,955-event test split; the noisy QSVC uses the smaller 500/2,000 split
described above and is included for comparison with the same caveat.

ERT achieves the highest overall accuracy (0.9857), precision (0.9723),
specificity (0.9715), and F1 score (0.9858) among all models evaluated,
reflecting its strong global classification performance on this dataset.
The ideal QSVC achieves the highest recall (0.999935), surpassing ERT
(0.999766), at the cost of a modestly higher false-positive rate: QSVC
specificity is 0.9550 compared to 0.9715 for ERT, and QSVC precision is
0.9569 compared to 0.9723 for ERT. The QSVC therefore represents a
different operating point on the precision--recall trade-off: it minimises
false negatives more aggressively than ERT, while accepting a somewhat
higher rate of background contamination.

This trade-off is of direct relevance in high-energy physics. In analyses
where preserving every signal event is critical — such as searches for rare
or exotic processes where signal yields are small — the QSVC's near-zero
false-negative rate may be preferable to the marginally higher overall
accuracy of ERT. Conversely, in applications requiring strong background
suppression at a fixed signal efficiency, ERT's higher precision and
specificity are advantageous. The two classifiers are therefore
complementary rather than strictly ordered, and the choice between them
should be guided by the physics requirements of the target analysis.

Among the classical SVM kernels, polynomial and RBF are competitive but
fall short of both QSVC and ERT across all metrics. The linear and sigmoid
kernels perform substantially worse, confirming that the centroid-based
feature space is strongly nonlinear and requires expressive nonlinear
similarity measures for effective discrimination. The noisy QSVC falls
below ERT, polynomial, and RBF on accuracy, precision, and specificity,
but outperforms the linear and sigmoid kernels on all three of those
metrics despite its 50-times-smaller test set, and notably retains a
recall (0.9951) that still exceeds every classical baseline except the
ideal QSVC and RBF kernel — underscoring that the recall-preserving
property of the quantum kernel, while attenuated, is not eliminated by
realistic hardware noise at the circuit depths considered here.

\begin{table*}[ht]
    \centering
    \resizebox{\linewidth}{!}{%
    \begin{tabular}{|c|c|c|c|c|c|c|c|}
    \hline
                 & ERT      & QSVC     & QSVC (noisy) & Linear   & Poly     & RBF      & Sigmoid \\ \hline
    Accuracy     & \cellcolor{green!20}0.985650 & \cellcolor{yellow!20}0.977450 & 0.897000 & 0.693856 & 0.970435 & 0.950072 & 0.479233 \\ \hline
    Precision    & \cellcolor{green!20}0.972307 & \cellcolor{yellow!20}0.956891 & 0.834299 & 0.638350 & 0.944311 & 0.909285 & 0.479300 \\ \hline
    Recall       & 0.999766 & \cellcolor{green!20}0.999935  & 0.995084 & 0.894875 & 0.999851 & \cellcolor{yellow!20}0.999930 & 0.477280 \\ \hline
    Specificity  & \cellcolor{green!20}0.971542 & \cellcolor{yellow!20}0.954979 & 0.795524 & 0.492718 & 0.941001 & 0.900183 & 0.481186 \\ \hline
    F1 score     & \cellcolor{green!20}0.985845 & \cellcolor{yellow!20}0.977940 & 0.907623 & 0.745153 & 0.971288 & 0.952456 & 0.478288 \\ \hline
    \end{tabular}%
    }
    \caption{Classification performance across all models. Green: best per row; yellow: second best. ERT leads on accuracy, precision, specificity, and F1; the ideal QSVC achieves the highest recall (13 false negatives across 401,955 test events), reflecting its bias toward signal retention over background suppression. The noisy QSVC (500 train / 2,000 test, FakeMumbaiV2 noise model) is evaluated on a substantially smaller test set than all other models and is not directly comparable in statistical precision, but retains a recall exceeding every classical baseline except RBF.}
    \label{tab:classification_results}
\end{table*}

\section{Model Analysis}
\label{sec:model_analysis}

\subsection{Receiver Operating Characteristic}

\begin{figure}[ht]
    \centering
    \begin{subfigure}{0.48\textwidth}
        \centering
        \includegraphics[width=\linewidth]{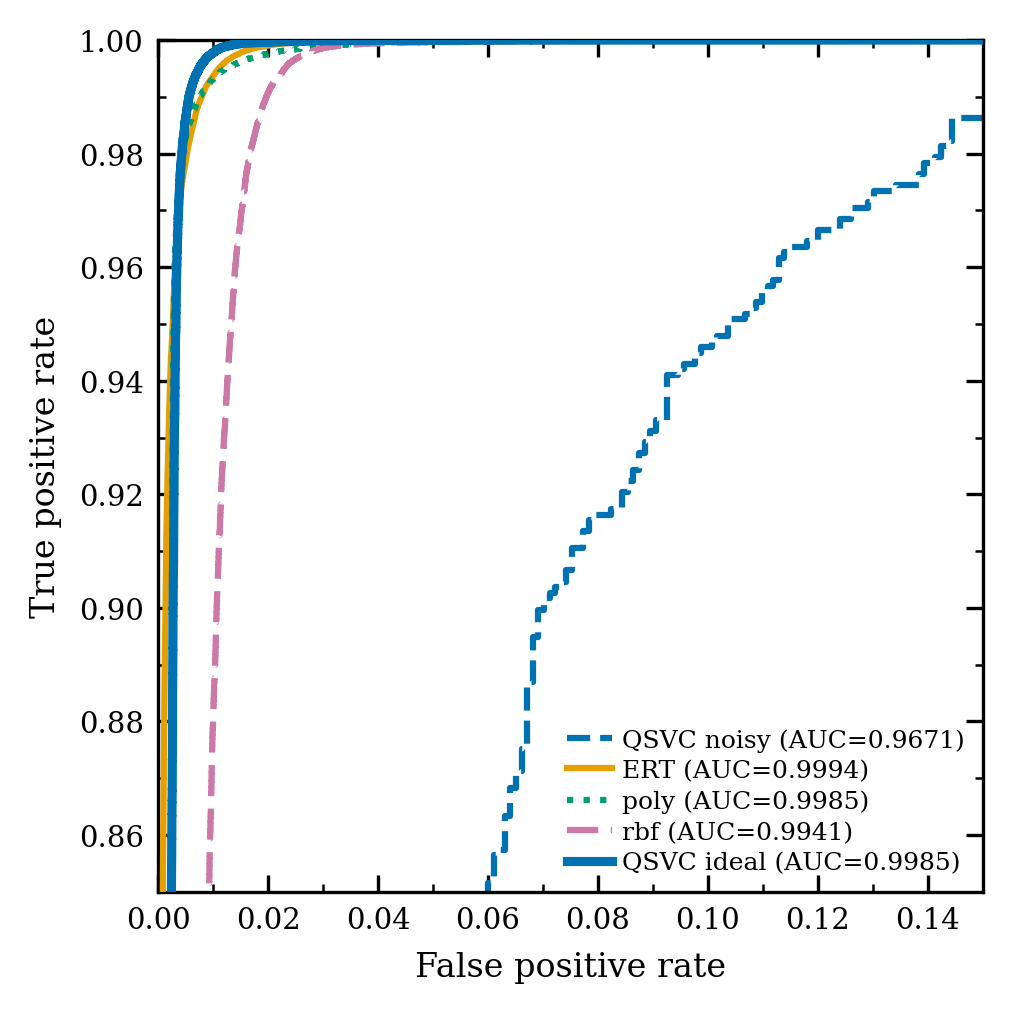}
        \caption{ROC curves comparing the ideal and noisy QSVC with
        classical baselines. The ideal QSVC achieves near-unity true
        positive rate at low false positive rates, with ERT attaining the
        highest global AUC; the noisy QSVC trails all ideal-simulation
        models but remains far above chance.}
        \label{fig:roc}
    \end{subfigure}
    \hfill
    \begin{subfigure}{0.48\textwidth}
        \centering
        \includegraphics[width=\linewidth]{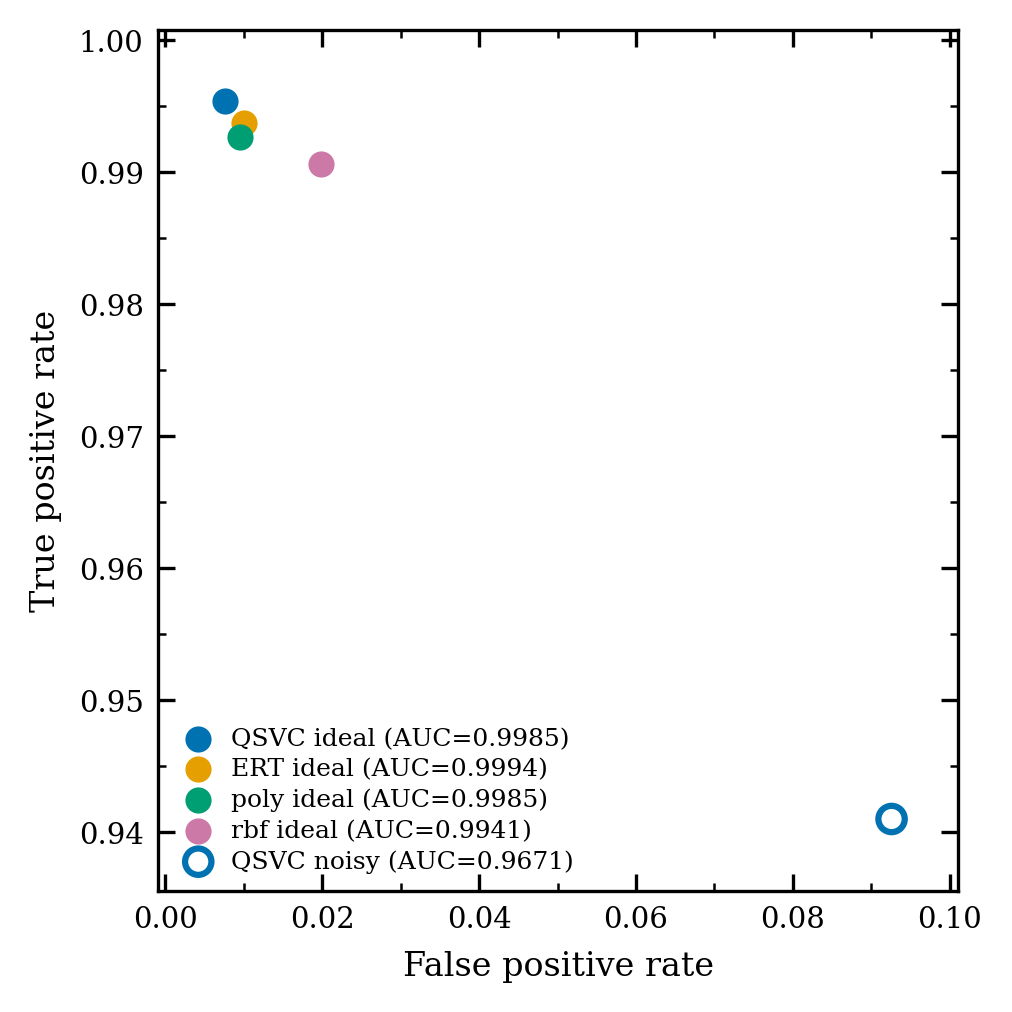}
        \caption{Optimal ROC operating points selected by minimising
        distance to the ideal classifier. The QSVC and ERT occupy
        distinct positions reflecting their different precision--recall
        trade-offs; the noisy QSVC operating point is well separated from
        the ideal-simulation cluster.}
        \label{fig:roc-points}
    \end{subfigure}
    \caption{ROC diagnostics comparing the QSVC with classical baselines.
    \textbf{Left:} Global ROC curves show that ERT achieves the highest
    AUC (0.9994), while the ideal QSVC (AUC~0.9985) exhibits a near-vertical
    rise from the origin consistent with its near-zero false-negative rate.
    The noisy QSVC (AUC~0.9671) shows a markedly slower rise, reflecting
    degraded background rejection under hardware noise.
    \textbf{Right:} Optimal operating points illustrate the complementary
    trade-offs between the two strongest ideal models: ERT maximises
    precision and specificity while the QSVC maximises recall; the noisy
    QSVC's operating point sits well outside this cluster, quantifying the
    performance cost of circuit-level decoherence.}
    \label{fig:roc_and_scores}
\end{figure}

Figure~\ref{fig:roc} compares the receiver operating characteristic (ROC)
curves for the ideal QSVC, the noisy QSVC, and all classical baselines. The
ideal QSVC achieves an AUC of 0.9985 and exhibits a near-vertical rise from
the origin, consistent with the confusion matrix result of only 13 false
negatives across more than 400,000 test events. Among all ideal-simulation
models, ERT achieves the highest global AUC (0.9994), confirming its strong
ranking performance on this dataset. The polynomial and RBF kernels are also
competitive, with AUC values of 0.9985 and 0.9941 respectively, confirming
that the centroid-based feature space is strongly nonlinear. The linear and
sigmoid kernels perform substantially worse (AUC 0.6735 and 0.4710),
consistent with the poor precision and recall reported in
Table~\ref{tab:classification_results}. The noisy QSVC (AUC~0.9671) trails
every ideal-simulation model shown, but remains far above the AUC~0.5
random-classifier baseline, indicating that meaningful class-discriminative
signal survives circuit-level decoherence at the noise levels modelled here.

Figure~\ref{fig:roc-points} shows the optimal operating point for each
model, selected by minimising the Euclidean distance to the ideal
classifier. The QSVC and ERT occupy distinct positions: the QSVC achieves
a higher true positive rate at a somewhat higher false positive rate,
while ERT achieves stronger background suppression at the cost of a
marginally lower signal efficiency. This is consistent with the
complementary precision--recall trade-off discussed in
Section~\ref{sec:evaluation}, and reinforces that neither model strictly
dominates the other across all operating regimes. The noisy QSVC's optimal
operating point is well separated from the ideal-simulation cluster,
providing a direct visual summary of the performance gap quantified
throughout this section.

\subsection{Significance Improvement Characteristic}

\begin{figure}[ht]
    \centering
    \begin{subfigure}{0.48\textwidth}
        \centering
        \includegraphics[width=\linewidth]{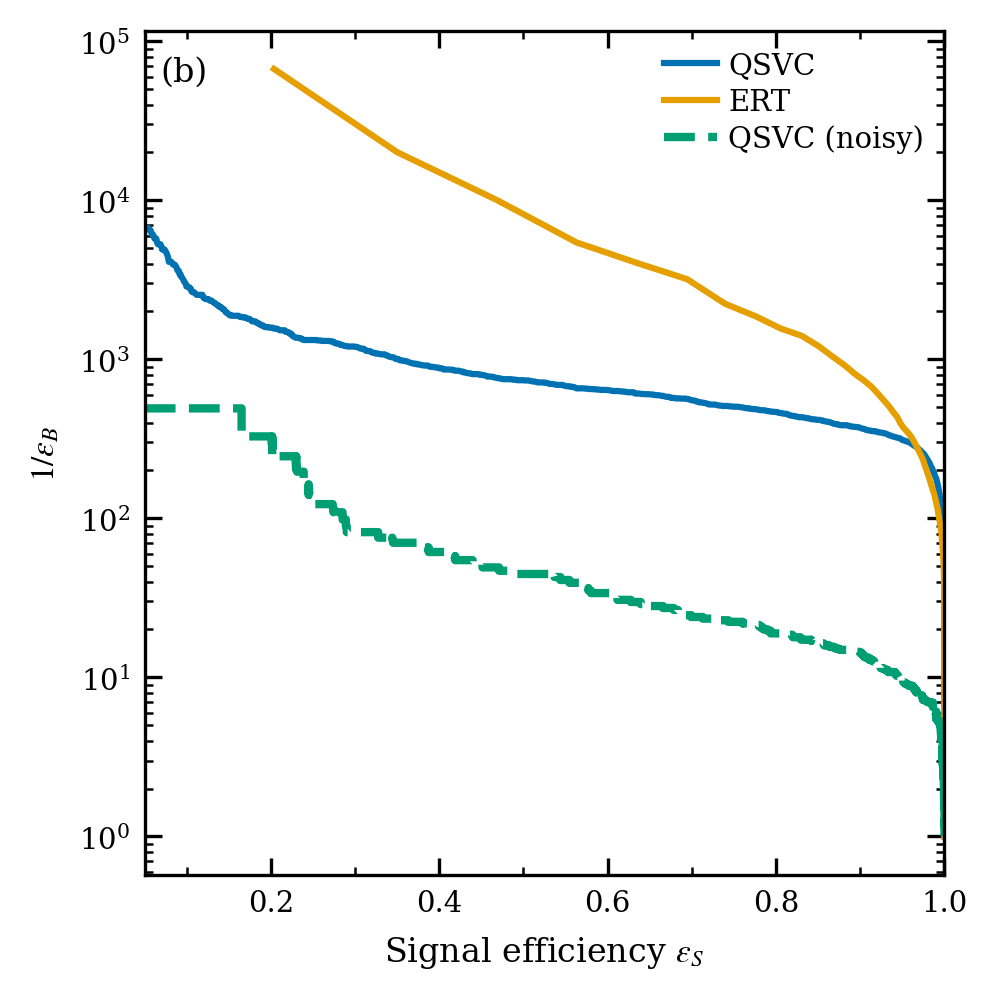}
        \caption{Background rejection $1/\epsilon_B$ as a function of
        signal efficiency $\epsilon_S$. ERT achieves stronger background
        rejection than the ideal QSVC across nearly the entire efficiency
        range, with the two curves converging only as $\epsilon_S \to 1$.
        The noisy QSVC trails both by two to three orders of magnitude
        across the full range.}
        \label{fig:sic-a}
    \end{subfigure}
    \hfill
    \begin{subfigure}{0.48\textwidth}
        \centering
        \includegraphics[width=\linewidth]{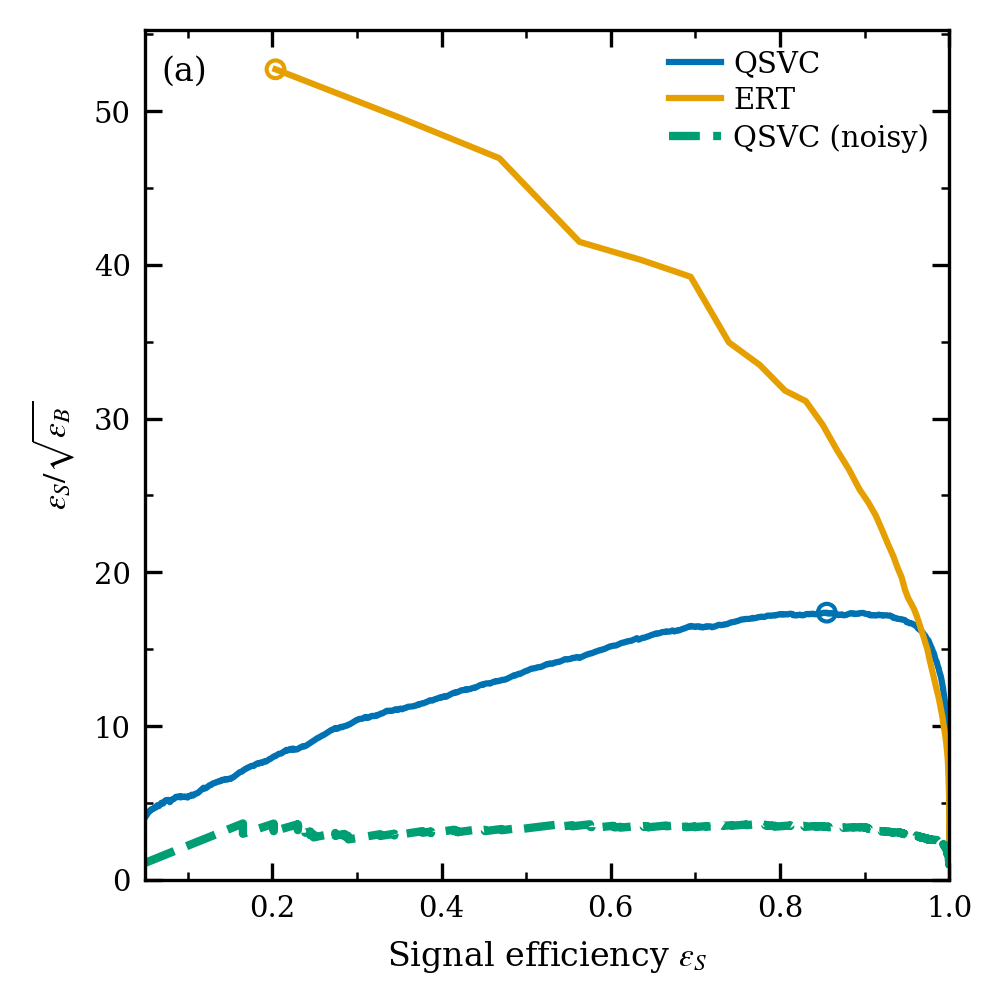}
        \caption{Significance improvement characteristic
        $\mathrm{SIC} = \epsilon_S / \sqrt{\epsilon_B}$ as a function of
        signal efficiency. The ideal QSVC exhibits a broad maximum at high
        $\epsilon_S$; ERT peaks sharply at low efficiency; the noisy QSVC
        plateaus at a substantially lower SIC across the full range.}
        \label{fig:sic-b}
    \end{subfigure}
    \caption{Physics-motivated performance diagnostics comparing the ideal
    QSVC, noisy QSVC, and ERT. \textbf{(a)} Background rejection curves
    show that ERT leads the ideal QSVC across nearly the entire efficiency
    range; the noisy QSVC's rejection curve sits well below both.
    \textbf{(b)} The ideal QSVC's SIC maximum occurs at much higher signal
    efficiency than ERT's, though at a lower absolute value; ERT's peak is
    higher but narrower and more sensitive to threshold choice. The noisy
    QSVC's flattened, low-amplitude SIC curve indicates that hardware
    noise degrades not only peak significance but also the
    threshold-tunability that makes a classifier operationally useful.}
    \label{fig:sic}
\end{figure}

The ROC analysis provides a global measure of ranking performance but does
not fully capture the physics relevance of different operating points. To
address this, Fig.~\ref{fig:sic} presents two complementary metrics
commonly used in high-energy physics: background rejection and the
significance improvement characteristic (SIC), both evaluated as functions
of signal efficiency.

The significance improvement characteristic (SIC), defined as $\varepsilon_S/\sqrt{\varepsilon_B}$,
and the background rejection $1/\varepsilon_B$ translate classifier performance from
information-theoretic metrics into the operational language of experimental nuclear physics.
In a counting experiment, the statistical significance of a signal above background scales as
$S/\sqrt{B}$, so applying a selection that retains signal efficiency $\varepsilon_S$ and
background efficiency $\varepsilon_B$ multiplies the baseline significance by exactly the SIC.
Since raw data collection improves significance only as $\sqrt{N}$, a classifier achieving
$\mathrm{SIC} = k$ provides the same statistical leverage as collecting $k^2$ times more data
with no selection applied. At its optimal threshold ($\varepsilon_S \approx 0.85$,
$\varepsilon_B \approx 0.0024$), the ideal QSVC retains 170,779 signal events from the
200,917-event test set while reducing the 201,038-event background to approximately 474 passing
events, yielding a peak $\mathrm{SIC} = 17.5$ and a data-collection equivalent of
$17.5^2 \approx 306{\times}$. ERT achieves a higher peak $\mathrm{SIC} = 53.1$ at
$\varepsilon_S \approx 0.20$, corresponding to a ${\sim}2{,}800{\times}$ equivalent, but does
so by discarding 80\% of signal events — an operating point that may be unsuitable for analyses
requiring large signal statistics, such as differential cross-section measurements. The noisy
QSVC plateaus at $\mathrm{SIC} \approx 3.5$ across the full efficiency range, equivalent to
${\sim}12{\times}$ more data, confirming that hardware noise at the current circuit depth does
not merely degrade accuracy but actively flattens the threshold-tunability that makes a
classifier operationally useful in an experimental analysis pipeline.

Figures~\ref{fig:sic-a} and~\ref{fig:sic-b} show that ERT achieves higher
absolute background rejection and SIC than the ideal QSVC across
essentially the entire efficiency range, including at the QSVC's own
optimal threshold, with the two curves converging only as
$\epsilon_S \to 1$. The distinguishing property of the ideal QSVC is
therefore not that it exceeds ERT at any shared operating point, but that
its own SIC-optimal threshold sits at $\epsilon_S \approx 0.85$ rather
than ERT's $\epsilon_S \approx 0.20$: an analysis that specifically
requires high signal statistics gains a more stable significance
enhancement from operating near the QSVC's own peak than from operating
ERT far from its optimum. The noisy QSVC trails both ideal-simulation
models by two to three orders of magnitude in background rejection and
shows the same SIC flattening noted above, directly reflecting the
degraded kernel geometry examined in Section~\ref{sec:data_analysis}.

ERT and the ideal QSVC are therefore not complementary in absolute
significance improvement — ERT achieves higher SIC and background
rejection at essentially every shared efficiency — but they are
complementary in where each model's own optimum lies: ERT's peak favours
highly selective, low-statistics selections, while the QSVC's peak favours
high-statistics analyses that require retaining most of the signal
population, consistent with its near-zero false-negative rate. The noisy
QSVC — though degraded relative to both — retains recall and background
rejection meaningfully above chance, motivating the noise- and
circuit-depth sensitivity analysis needed before deployment on genuine
NISQ hardware.

\section{Data Analysis}
\label{sec:data_analysis}

\subsection{PCA and t-SNE Embeddings of Quantum Statevectors}

\begin{figure}[ht]
    \centering
    \begin{subfigure}{0.48\textwidth}
        \centering
        \includegraphics[width=\linewidth]{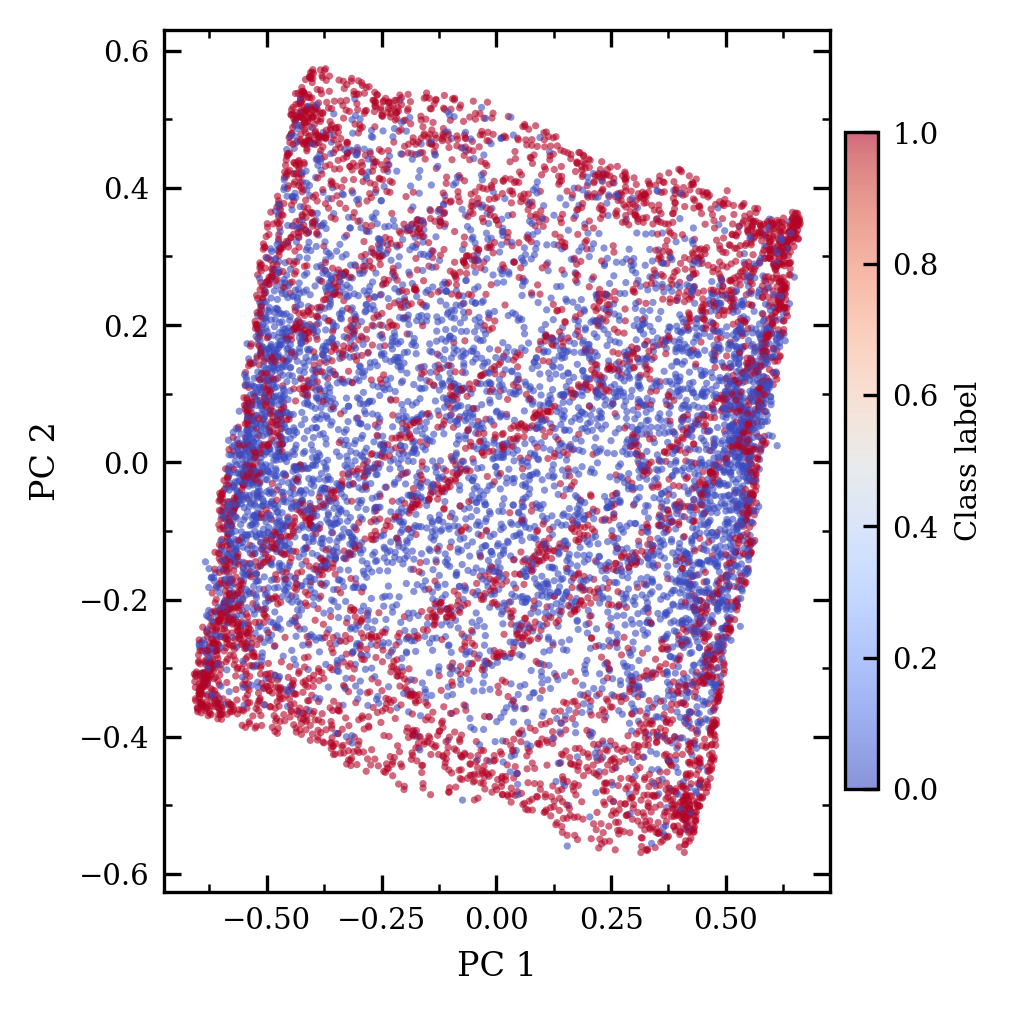}
        \caption{PCA projection of the encoded quantum statevectors. 
        The overlap between classes reflects the limitation of linear projections 
        rather than the separability achievable in the full Hilbert space.}
        \label{fig:pca}
    \end{subfigure}
    \hfill
    \begin{subfigure}{0.48\textwidth}
        \centering
        \includegraphics[width=\linewidth]{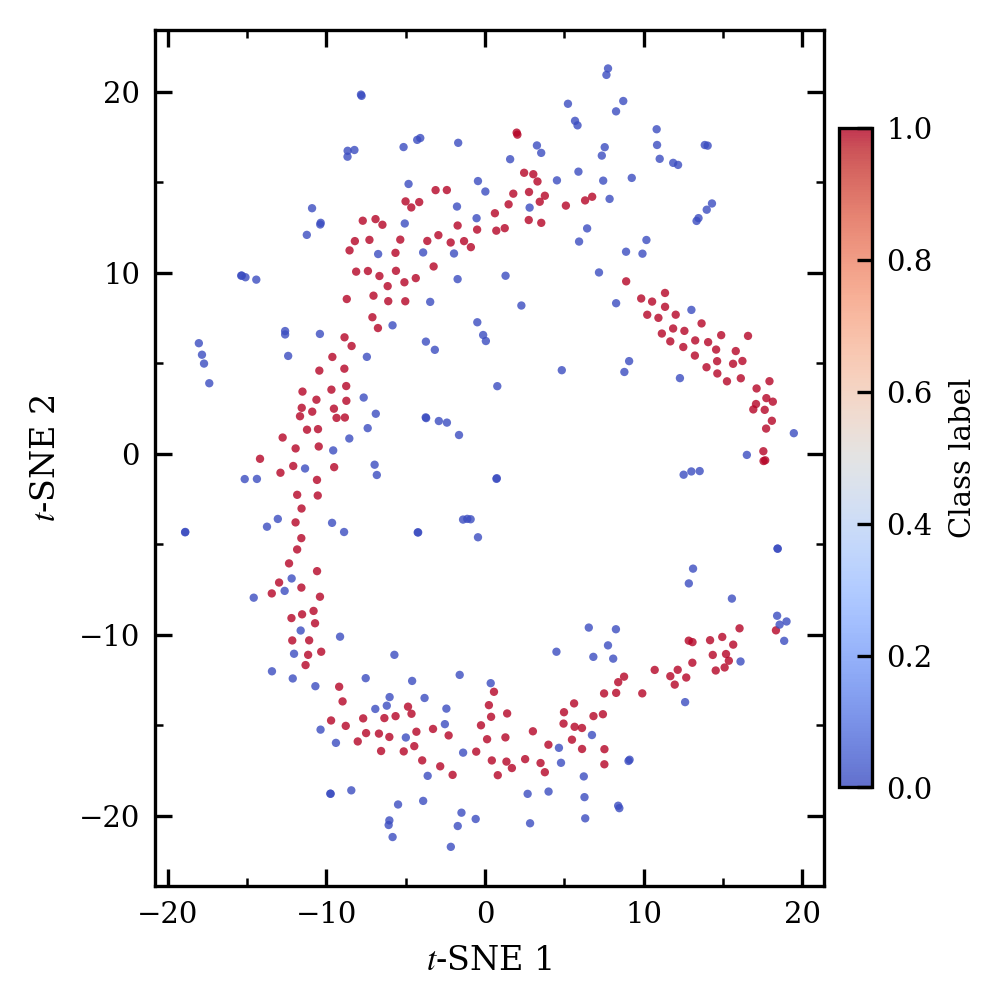}
        \caption{t-SNE visualization constructed from quantum-kernel distances. 
        The embedding reveals a curved, structured manifold with locally distinct 
        class neighborhoods shaped by the quantum feature map.}
        \label{fig:tsne}
    \end{subfigure}

    \caption{
    Visualizations of the quantum feature space produced by the ZZFeatureMap. 
    \textbf{Left:} PCA shows that although the embedding is highly expressive, 
    its nonlinear structure cannot be captured by linear projections. 
    \textbf{Right:} t-SNE reveals meaningful geometric organization, with 
    signal and background forming intertwined but locally distinguishable regions. 
    Together, these projections illustrate that the quantum kernel induces 
    a smooth, nonlinear manifold structure that enables the QSVC to construct 
    a sharp decision boundary in the full Hilbert space.
    }
    \label{fig:pca_tsne}
\end{figure}

Figure~\ref{fig:pca} shows the first two principal components of the 64-dimensional complex statevectors generated by the ZZFeatureMap under ideal simulation; as with the entanglement entropy and fidelity analyses in Section~\ref{sec:embedding_geometry}, this requires explicit statevectors and is not repeated for the noisy compute-uncompute kernel. As expected for a linear projection of a nonlinear kernel, the two classes largely overlap.

The encoded states nonetheless lie on a smooth, bounded manifold rather than filling the plane randomly — a first indication of the structured, non-maximal entanglement examined quantitatively in Section~\ref{sec:embedding_geometry}.

Figure~\ref{fig:tsne} shows a t-SNE embedding constructed from the same quantum-kernel distances; unlike PCA, t-SNE preserves local neighborhoods and is therefore sensitive to nonlinear structure. The resulting manifold is curved and continuous, with the two classes occupying intertwined but locally distinguishable regions, reflecting the intra-class similarity and inter-class separation quantified in the fidelity analysis below.

Taken together, Figures~\ref{fig:pca} and~\ref{fig:tsne} provide visual
corroboration for the quantitative geometric analyses that follow, rather
than standing as an independent result. The t-SNE embedding of
Figure~\ref{fig:tsne} is constructed directly from the same pairwise
quantum-kernel distances characterised statistically in
Figure~\ref{fig:fidelity} and the kernel quadrant box plot of
Figure~\ref{fig:qkernel_boxplot}: the locally distinguishable but
intertwined class neighborhoods visible in the t-SNE manifold are the
direct visual manifestation of the same intra-class fidelity bias and
inter-class near-orthogonality quantified numerically in those figures.
Similarly, the smooth, bounded manifold observed in the PCA projection of
Figure~\ref{fig:pca} — rather than a diffuse, space-filling point cloud —
is consistent with the moderate, non-maximal entanglement entropy reported
in Figure~\ref{fig:entanglement}: a maximally entangled, Haar-random
encoding would, by concentration of measure, be expected to populate any
low-dimensional projection more
uniformly~\cite{McClean2018,Holmes2022}, whereas the observed structured, bounded manifold instead
reflects the same intermediate, data-dependent entanglement regime
measured directly in Section~\ref{sec:embedding_geometry}. The predictable
manifold and local clustering visible here therefore support, and are
supported by, the quantitative results of Figures~\ref{fig:entanglement},
\ref{fig:fidelity}, and~\ref{fig:qkernel_boxplot}: the qualitative
geometry visualised in this subsection and the quantitative statistics
reported in the following two are two independent views of the same
underlying Hilbert-space structure, and their mutual consistency is
itself evidence that neither is an artefact of the specific analysis
method used to produce it.

\subsection{Quantum Embedding Geometry: Entanglement and Fidelity Structure}
\label{sec:embedding_geometry}

\begin{figure}[ht]
    \centering
    \begin{subfigure}{0.48\textwidth}
        \centering
        \includegraphics[width=\linewidth]{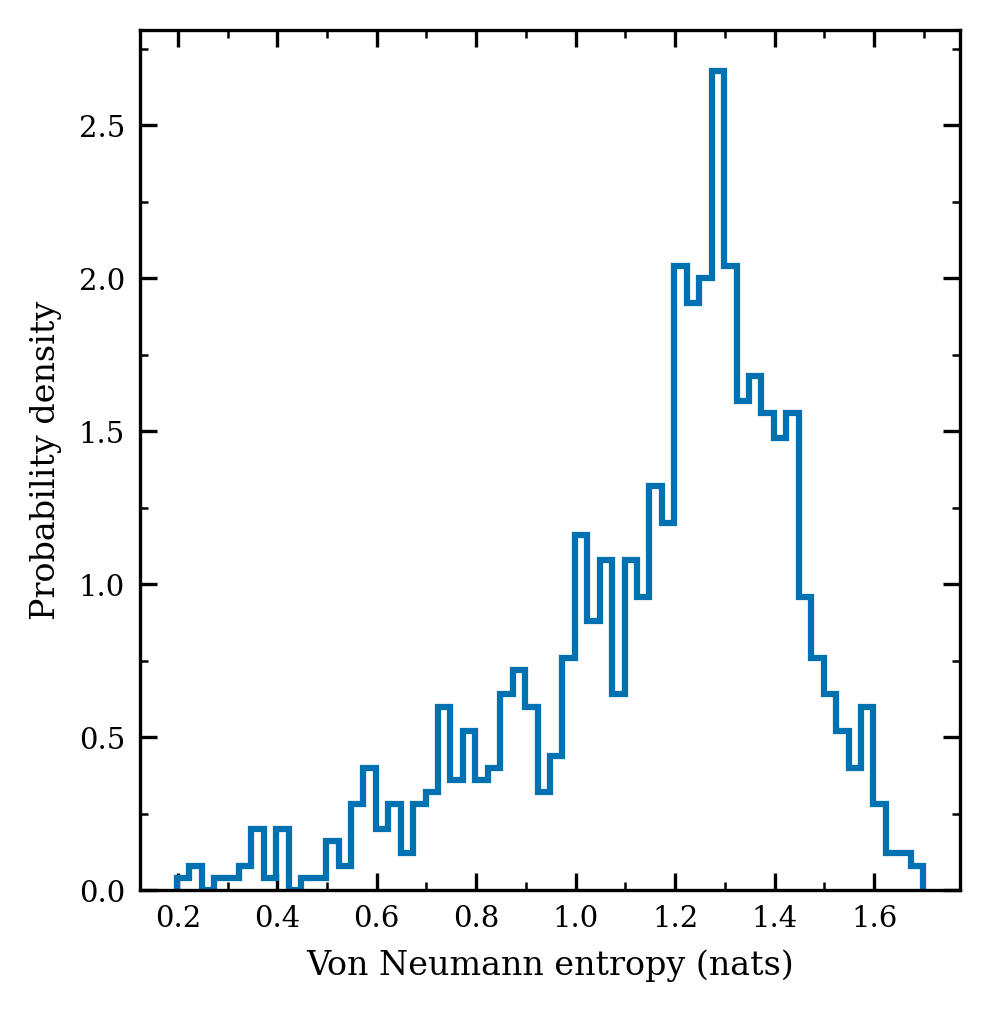}
        \caption{Distribution of bipartition Von Neumann entanglement
        entropy (3-qubit vs.\ 3-qubit bipartition) for the encoded
        statevectors. The ZZFeatureMap produces moderately high but
        variable entanglement across the dataset.}
        \label{fig:entanglement}
    \end{subfigure}
    \hfill
    \begin{subfigure}{0.48\textwidth}
        \centering
        \includegraphics[width=\linewidth]{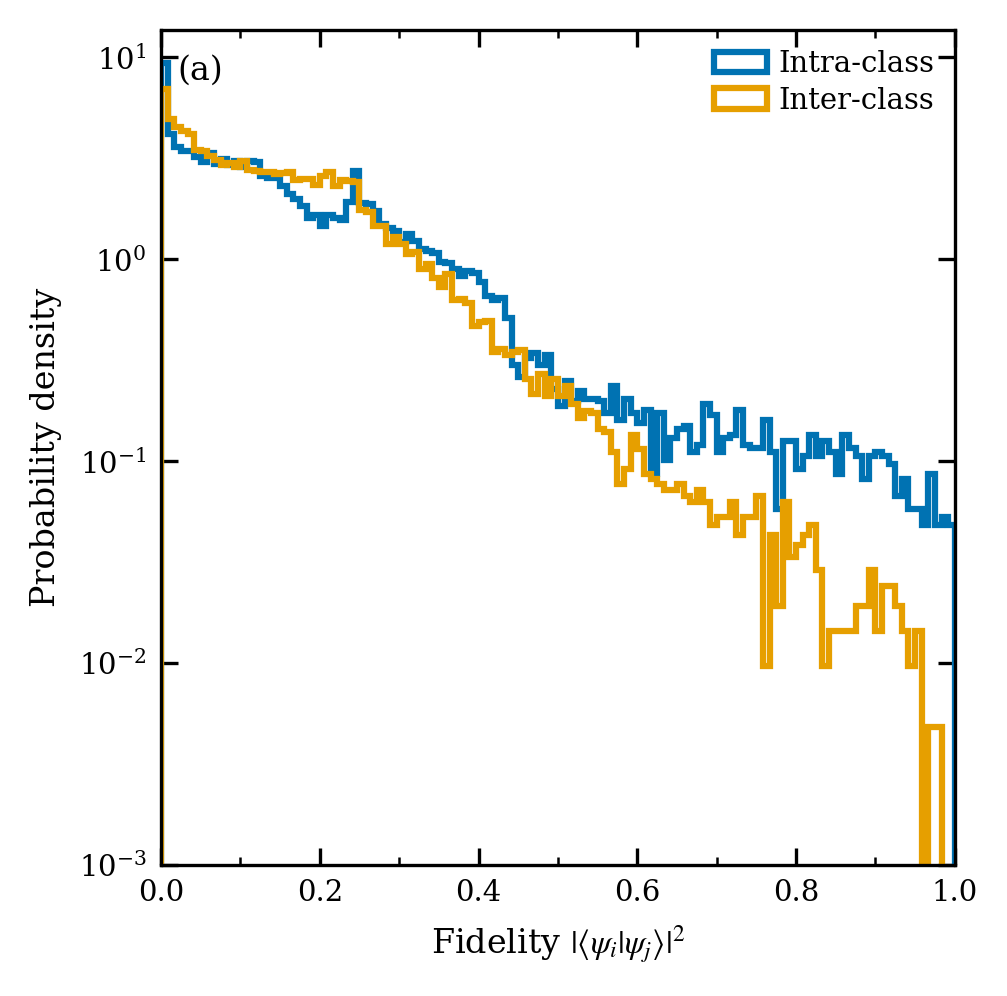}
        \caption{Pairwise fidelity distributions for intra-class and
        inter-class statevector pairs under ideal simulation. Inter-class
        pairs concentrate near zero fidelity; intra-class pairs retain a
        heavier tail at moderate overlap.}
        \label{fig:fidelity}
    \end{subfigure}
    \caption{Quantum-state structure induced by the ZZFeatureMap under
    ideal simulation. \textbf{Left:} Entanglement entropy characterises
    the expressivity of the static feature-map encoding. \textbf{Right:}
    Fidelity distributions reveal that inter-class states tend toward
    near-orthogonality while intra-class states retain coherent local
    neighbourhoods, consistent with the ideal QSVC's decision-boundary
    geometry. A quantitative, ideal-versus-noisy comparison of this
    fidelity structure is presented in Section~\ref{sec:qkernel_boxplot}.}
    \label{fig:entropy_and_fidelity}
\end{figure}

Figure~\ref{fig:entanglement} shows the distribution of Von Neumann
entanglement entropy for a balanced 3-qubit vs.\ 3-qubit bipartition of the
encoded statevectors. Entropy values span a wide range (approximately
0.2--1.7 nats), indicating that the ZZFeatureMap produces a diverse family
of states rather than collapsing to trivially unentangled or maximally
random extremes. This analysis is restricted to ideal statevectors for two
reasons. Practically, the noisy compute-uncompute protocol
(Section~\ref{sec:noisy_sim}) never constructs an explicit per-sample
statevector — it only measures pairwise circuit outcomes used directly as
kernel entries — so no state exists on which the SVD-based entropy
calculation could be evaluated without a separate density-matrix
simulation. More fundamentally, even that pass would not yield a
comparable quantity: under noise each state is mixed rather than pure,
and for a mixed state the reduced-subsystem entropy conflates entanglement
with noise-induced decoherence — a fully depolarized, unentangled state
can still show high reduced-subsystem entropy purely from mixedness. We
therefore report entanglement entropy for the ideal case only; the
noise-driven compression of class-dependent kernel structure is instead
characterised directly via the quadrant-resolved kernel fidelity analysis
(Section~\ref{sec:qkernel_boxplot}) and the noisy kernel matrices
(Section~\ref{sec:noisy_sim}), which carry no such interpretive ambiguity.

Most samples fall in the moderately entangled range (1.2--1.5 nats),
consistent with the fully entangling circuit structure. Computed on the
static, untrained feature-map encoding, this entropy characterises the
geometry of the embedding itself rather than learned classifier behaviour:
the moderate, data-dependent values show that the ZZFeatureMap produces an
input-sensitive embedding whose expressivity varies with track geometry,
with the low-entropy tail corresponding to events in simpler regions of
the input space.

Figure~\ref{fig:fidelity} shows the pairwise state fidelities for
intra-class and inter-class sample pairs under ideal simulation (50,000
randomly drawn pairs per category). Both distributions skew toward low
fidelities, confirming that the ZZFeatureMap spreads samples widely
throughout the 64-dimensional Hilbert space. Inter-class fidelities
concentrate closer to zero than intra-class fidelities, indicating that
signal and background states tend toward near-orthogonality, while
intra-class pairs retain a heavier tail at moderate fidelities
(0.2--0.6), forming coherent local neighbourhoods. Since the QSVC kernel
is defined directly in terms of pairwise fidelity, this structure
contributes directly to classification — low inter-class fidelity
increases effective class separation, moderate intra-class fidelity
stabilises the SVM margin — providing a geometric explanation for the
ideal QSVC's high recall and broad SIC plateau. Whether this structure
survives under hardware noise is examined quantitatively in the
quadrant-resolved kernel fidelity analysis of
Section~\ref{sec:qkernel_boxplot}, which explains the noisy QSVC's
retained-but-degraded performance documented above.

The degree of inter-class orthogonality observed here is a known property
of high-frequency ZZFeatureMaps on normalised input domains, arising as a
mathematical consequence of the encoding rather than necessarily
reflecting class-aware structure learned during
training~\cite{Schnabel_2025}; the fidelity distributions are
therefore best read as characterising embedding geometry rather than
direct evidence of a quantum advantage.

\subsection{Quantum Kernel Fidelity Distributions}
\label{sec:qkernel_boxplot}

\begin{figure}[ht]
    \centering
    \includegraphics[width=0.85\linewidth]{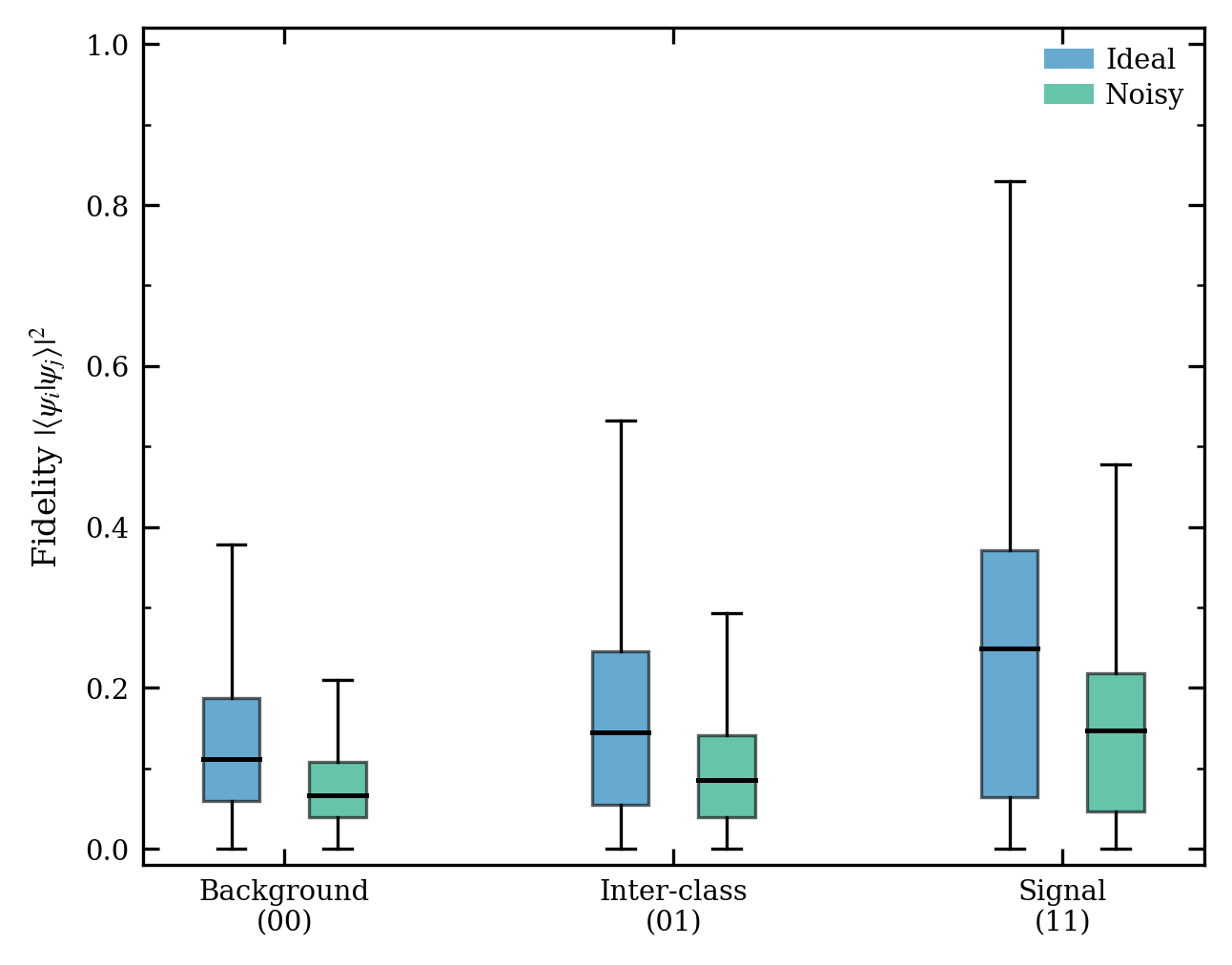}
    \caption{
    Fidelity distributions $|\langle\psi_i|\psi_j\rangle|^2$ between encoded
    samples, grouped by class-pair category — background–background (00),
    inter-class (01), and signal–signal (11) — for the ideal (blue) and
    noisy (green) quantum kernels. Boxes show the median and interquartile
    range; whiskers extend to $1.5\times$IQR. Diagonal self-fidelity entries
    are excluded from all categories.}
    \label{fig:qkernel_boxplot}
\end{figure}

Figure~\ref{fig:qkernel_boxplot} quantifies the class-dependent structure of
the quantum kernel visible qualitatively in the fidelity histograms. Under
ideal simulation, the three categories are clearly separated in median
fidelity, with signal–signal pairs (11) exhibiting the highest median
similarity ($\approx 0.25$), inter-class pairs (01) intermediate
($\approx 0.15$), and background–background pairs (00) the lowest
($\approx 0.11$). This ordering indicates that signal events occupy a
comparatively compact region of the encoded Hilbert space, while background
events are more mutually dissimilar — providing the geometric basis for the
QSVC decision boundary.

Under FakeMumbaiV2 noise, all three distributions are compressed toward
lower fidelity and the interquartile ranges narrow substantially, consistent
with noise-driven concentration of the encoded states toward the maximally
mixed state. Critically, however, the relative ordering
$00 < 01 < 11$ is preserved: the noisy medians follow the same class-dependent
structure as the ideal kernel, only attenuated rather than erased. This
explains why the noisy QSVC retains meaningful classification power
(AUC $= 0.9671$) rather than degrading to chance — sufficient
class-discriminative signal survives circuit-level decoherence for the SVM
to recover a usable, if weaker, margin. The magnitude of this attenuation is
the direct kernel-level signature of the performance gap quantified in the
ROC and SIC comparisons of Section~\ref{sec:model_analysis}.

\section{Discussion}
\label{sec:discussion}

The results presented in this work demonstrate that quantum kernel methods can provide a meaningful advantage for track–based event classification when applied to compressed detector features, under ideal simulation. The ideal QSVC achieves the highest recall of any model evaluated and remains within one to two percentage points of the best classical model (ERT) on accuracy, precision, specificity, and F1 score, while exhibiting a dramatically lower false negative rate. This behavior is especially valuable in high-energy physics, where missed signal events correspond directly to lost physics opportunities and cannot be recovered downstream. Critically, this work also characterises how much of that advantage survives under a realistic hardware noise model, providing — to our knowledge — one of the more direct ideal-versus-noisy comparisons of quantum-kernel classification performance on real detector data reported to date.

The geometric analyses of the quantum embedding offer insight into why the QSVC performs so well under ideal simulation, and why its performance degrades in a specific, structured way under noise. The ZZFeatureMap generates quantum states with moderately high but data-dependent entanglement, avoiding both trivial separability and fully random embeddings. Fidelity and distance histograms show that the feature map spreads samples widely throughout the Hilbert space while preserving coherent intra-class neighborhoods. The resulting quantum kernel matrices are highly expressive, with most sample pairs nearly orthogonal yet containing localized regions of higher similarity that reflect meaningful detector-level structure. This geometry enables the SVM to form a sharp, high-margin decision boundary, consistent with the near-perfect recall achieved in testing.

The quadrant-resolved fidelity analysis (Section~\ref{sec:qkernel_boxplot}) shows that this class-dependent kernel structure is not destroyed by hardware noise, but rather uniformly compressed: the ordering of median fidelity across background-background, inter-class, and signal-signal pairs is preserved under the FakeMumbaiV2 noise model, while the absolute fidelity values and their spread both shrink substantially, consistent with noise-driven concentration of the encoded states toward the maximally mixed state. This finding directly explains the noisy QSVC's behaviour: rather than collapsing to a random classifier, it retains a meaningfully weaker version of the ideal kernel's discriminative geometry, which manifests as a noisy AUC of 0.9671 (versus 0.9985 ideal) and a recall of 99.51\% (versus 99.99\% ideal) that remains higher than every classical baseline except RBF. The SIC analysis (Section~\ref{sec:model_analysis}) shows the cost of this compression most clearly in the loss of threshold-tunability: the noisy QSVC's significance-improvement curve is not merely lower in amplitude than the ideal curve, it is nearly flat, meaning that unlike the ideal QSVC an analyst has little ability to trade signal efficiency against background rejection by adjusting the classification threshold. This is arguably a more serious practical limitation than the AUC or accuracy drop alone, since operational deployment of a classifier in a physics analysis pipeline typically depends on the ability to tune the operating point to the requirements of a specific measurement.

The classical models provide an informative comparison. While nonlinear kernels such as polynomial and RBF approach the ideal QSVC in AUC, they do not reproduce the same degree of inter-class orthogonality or the tight intra-class clustering observed in the quantum embedding, and their false negative rates remain correspondingly higher. Theoretical work suggests that quantum kernels can, in principle, induce effectively very high-order polynomial structure through circuit-based feature maps, yielding a richer implicit feature space than a fixed-degree polynomial kernel~\cite{Havlicek2019}; whether this specific mechanism explains the performance gap observed here is plausible but was not directly tested in this work. Linear and sigmoid kernels perform significantly worse, reinforcing the conclusion that the underlying feature space is strongly nonlinear and highly structured.

Although ERT achieves the highest overall ROC AUC, the ideal QSVC exhibits a slightly superior optimal operating point in terms of proximity to the ideal classifier, and its SIC-optimal threshold sits at a substantially higher signal efficiency than ERT's (Section~\ref{sec:model_analysis}). This indicates that while ERT provides stronger global ranking performance and higher absolute significance improvement, the QSVC's own optimum is better suited to analyses that specifically require high signal statistics. The noisy QSVC does not compete with either model on global ranking metrics, but its retained recall advantage over all classical baselines except RBF suggests that even at current NISQ-era noise levels, the qualitative signal-retention behaviour of the quantum kernel is not entirely an artefact of noiseless simulation.

From a computational standpoint, the primary limitation of this approach is the cost of statevector simulation and, more severely, noisy circuit-level kernel construction. Once ideal quantum statevectors are generated, kernel evaluation is efficient and reusable, allowing for rapid experimentation with SVM configurations — the entire ideal QSVC path (statevector generation through test inference) required under 22 minutes of wall-clock time on our 5-node, 200-core allocation. By contrast, the noisy kernel study, evaluated on a subsample more than 50 times smaller than the ideal test set, still required nearly 80 minutes, reflecting the fundamentally different cost structure of measurement-based circuit evaluation versus a cached statevector inner product. Distributing the noisy test-kernel computation across MPI ranks (Section~\ref{sec:hpc}) was necessary to make even this reduced-scale noisy study tractable; scaling the noisy study to match the full 401,955-event ideal test set would require either substantially greater compute allocation, a reduction in shot count per kernel entry (at the cost of increased statistical noise in the kernel itself), or an approximate kernel evaluation strategy. We note also that the classical SVM baselines evaluated here — using Scikit-Learn's general-purpose, non-precomputed-kernel solver — were themselves the single most expensive component of the full pipeline, consuming 69\% of total wall-clock time despite requiring no quantum simulation at all; this is a reminder that "classical" does not imply "cheap," and that fair wall-clock comparisons between classical and quantum-kernel methods depend heavily on implementation choices on both sides. This workflow highlights a practical intermediate regime in which quantum-inspired methods — using quantum feature maps simulated classically, with noise characterisation via calibrated noise models — can already provide benefits and honest performance bounds even before fault-tolerant hardware becomes widely accessible.

Overall, these findings indicate that quantum-enhanced kernel methods can offer a distinct advantage for classification tasks where (1) high recall is critical, (2) the data possesses nonlinear correlations not easily captured by classical kernels, and (3) moderately expressive quantum feature maps can be constructed — while also making clear that this advantage is presently contingent on ideal or near-ideal simulation conditions, and degrades in a specific, quantifiable way as circuit-level noise is introduced. Chief among the directions for future work is evaluation of the noisy QSVC pipeline on physical quantum hardware rather than simulated noise models, which we regard as necessary to confirm that the noise-driven degradation characterised in this work is representative of genuine device behaviour rather than an artefact of the FakeMumbaiV2 calibration snapshot. Beyond hardware validation, future research may explore hardware-efficient encodings, error-mitigated kernel evaluation, hybrid classical–quantum architectures, and applications to higher-dimensional detector representations. Applying these techniques to online triggering or real-time streaming environments may further reveal scenarios where the geometric properties of quantum kernels yield unique performance benefits, provided the noise sensitivity documented here can be addressed through improved hardware fidelity, error mitigation, or noise-aware training strategies.

\subsection*{Alignment and Gaps Summary}

This work directly addresses one of the central open questions in applying quantum kernel methods to high-energy physics — the gap between simulator-based performance claims and behaviour under realistic hardware noise — by providing a quantitative, same-dataset comparison between ideal and noisy quantum kernel classification. Several related gaps nonetheless remain:

\begin{enumerate}
    \item \textbf{Simulated Noise Models versus Physical Hardware.}
    A central limitation of this work is that the noisy results reported throughout this paper are obtained from a calibrated noise model (FakeMumbaiV2) applied within classical simulation, and do not reflect execution on physical quantum hardware. While such noise models are constructed from real device calibration data and are widely used as a proxy for hardware behaviour, they do not capture all sources of hardware-specific error — including drift over time, crosstalk beyond the modelled two-qubit error rates, and correlated noise processes — and may therefore over- or under-state the true performance degradation a physical device would exhibit. Additionally, the noisy study in this work was evaluated on a substantially smaller sample (500 train / 2,000 test) than the ideal study (100,488 train / 401,955 test), due to the computational cost of circuit-level measurement simulation documented in Section~\ref{sec:hpc}. We regard evaluation on physical quantum hardware — as opposed to simulated noise models — as the most important open validation step for this work, and we plan to pursue direct hardware execution of the noisy QSVC kernel construction and classification pipeline as future work.

    \item \textbf{Fake Rate Stability.}
    Studies in track reconstruction, particularly at DESY, indicate that quantum approaches can
    exhibit elevated fake rates relative to mature classical algorithms. The noisy QSVC's reduced
    specificity (0.7955, versus 0.9550 under ideal simulation) documented in
    Section~\ref{sec:noisy_eval} is broadly consistent with this pattern. Addressing this
    instability is essential before quantum models can be deployed in real-time or trigger-level
    systems where reliability is critical.

    \item \textbf{Automatic Architecture Search.}
    Most current approaches, including this work, rely on fixed feature maps (e.g., ZZFeatureMap or first-order
    encodings). Emerging research at Brookhaven National Laboratory suggests a transition
    toward adaptive, data-driven circuit design through techniques such as reinforcement
    learning and distributed quantum sensing. These approaches aim to develop
    ``AI-native'' quantum architectures capable of adapting to detector geometry and hardware
    noise characteristics — potentially including adapting circuit depth or structure specifically
    to preserve the class-dependent kernel geometry identified in Section~\ref{sec:qkernel_boxplot}
    under noise.
\end{enumerate}

\section{Conclusion}

In this work we have demonstrated that a quantum support vector classifier, built from a fully entangling ZZFeatureMap, can achieve highly effective particle–track discrimination using compact centroid-based detector features. Under ideal simulation, the QSVC consistently outperforms all four classical SVM kernels evaluated and remains within one to two percentage points of the best published classical model (ERT) on accuracy, precision, specificity, and F1 score, while achieving the highest recall of any model and delivering an exceptionally low false–negative rate. This is of particular importance in high-energy physics, where missed signal events correspond directly to lost physics sensitivity.

The accompanying geometric analyses reveal that the quantum feature map produces a structured, expressive embedding in which inter-class states become nearly orthogonal while intra-class neighborhoods remain coherent. This combination of global expressivity and local structure enables the ideal QSVC to form a sharp, high-margin decision boundary that classical kernels struggle to reproduce. Together, these findings provide evidence that quantum-enhanced kernels can capture nonlinear correlations in compact detector data that are difficult to model with conventional methods.

Beyond the ideal-simulation results, we have also characterised how this performance degrades under a realistic hardware noise model (FakeMumbaiV2). The noisy QSVC's AUC falls from 0.9985 to 0.9671 and its peak significance improvement characteristic falls from 17.5 to approximately 3.5 — a significant reduction in physics-relevant statistical leverage. Notably, however, the noisy QSVC's signal recall (99.51\%) remains close to its ideal-simulation value (99.99\%) and above every classical baseline except RBF, indicating that the quantum kernel's central practical advantage — near-complete signal retention — is attenuated but not eliminated by circuit-level decoherence at current NISQ-era noise levels. The quadrant-resolved kernel fidelity analysis shows why: hardware noise compresses the class-dependent fidelity structure of the quantum kernel toward the maximally mixed state without erasing its relative ordering, leaving a weaker but still meaningfully structured decision geometry.

Statevector simulation and, more substantially, noisy circuit-level kernel evaluation remain the dominant computational costs of this workflow. The ability to cache ideal quantum embeddings and reuse kernel evaluations makes the ideal-simulation workflow practical for large classical datasets, with the full ideal QSVC pipeline completing in under 22 minutes on a 5-node, 200-core HPC allocation. To our knowledge, the 502,443-event dataset and exactly constructed $100{,}488\times100{,}488$ ideal kernel matrix evaluated in this work represent one of the largest datasets used in a published quantum-kernel classification study to date, more than an order of magnitude larger than the largest comparable study we identified in the literature~\cite{Wu2021}, made tractable specifically through the multi-node HPC architecture described above. The noisy study, by contrast, required distributing compute-uncompute circuit evaluation across all available MPI ranks to remain tractable even at a sample size more than 50 times smaller than the ideal test set — underscoring that noisy kernel evaluation, not classical simulation of the ideal case, is the binding constraint on scaling this approach further. As quantum hardware advances, hardware-efficient encodings, error mitigation, and hybrid quantum–classical strategies may reduce this overhead and narrow the ideal-to-noisy performance gap documented here, enabling deployment in real-time or near–real-time analysis chains.

Overall, this study demonstrates that quantum kernel methods already offer tangible benefits in realistic physics classification settings under ideal simulation, and that a meaningful — if reduced — fraction of that benefit persists under a calibrated hardware noise model. Their strong recall, robustness, and rich geometric structure suggest a promising role for QML in future high-energy physics analyses, particularly in regimes where maximizing signal retention is essential, provided the noise sensitivity quantified in this work continues to be addressed through improved hardware fidelity, error mitigation, and noise-aware circuit design.

\section*{Data Availability}

The data that supports the findings of this study is from the Thomas Jefferson National Accelerator Facility (JLab).
Restrictions apply to the availability of this data.
Data is available from the authors upon reasonable request and with permission from JLab.

\bibliographystyle{elsarticle-num} 
\bibliography{bibliography.bib}






\end{document}